\documentclass[11pt]{article} 

\usepackage[utf8]{inputenc} 
\usepackage[english]{babel}
\usepackage{csquotes}

\usepackage{geometry} 
\usepackage{graphicx} 
\usepackage[outdir=./]{epstopdf}

\usepackage[parfill]{parskip} 

\usepackage{amsmath,amssymb} 
  
\usepackage{booktabs} 
\usepackage{tabularray} 
\usepackage{tablefootnote}
\usepackage{float} 
\floatstyle{plaintop}
\restylefloat{table}

\usepackage{caption}  
\usepackage{multirow} 
\usepackage{array} 
\usepackage{etoolbox} 
\usepackage{paralist} 
\usepackage{verbatim} 
\usepackage{subfig} 
\usepackage[colorlinks=true,
			citecolor=blue,
			linkcolor=blue,
            breaklinks=true,
 			urlcolor=blue]{hyperref} 

\usepackage{tikz} 
\usepackage{xcolor,color}
\usetikzlibrary{fit}

\usepackage[title]{appendix}

\usepackage{authblk}  

\usepackage{setspace} 
\usepackage[capitalise,nameinlink]{cleveref}
\crefname{supp}{Supplement}{Supplements}

\usepackage{fancyhdr} 
\usepackage[
authordate,
giveninits=true,
uniquename=false,
backend=biber,
natbib=true,
doi=only,
isbn=false,
url=false,
maxcitenames=2,
noibid
]{biblatex-chicago}

\NewBibliographyString{available}
\NewBibliographyString{urlwhen}

\DefineBibliographyStrings{english}{%
  available = {Available at},
}

\DefineBibliographyStrings{english}{%
  urlwhen = {accessed},
}

\DeclareFieldFormat{url}{\bibstring{available}\addcolon\space\url{#1}}
\DeclareFieldFormat{urldate}{\mkbibparens{\bibstring{urlwhen}\addcolon\space{#1}}}

\DeclareCiteCommand{\citeyear}
    {}
    {\bibhyperref{\printdate}}
    {\multicitedelim}
    {}

\usepackage[hyphenbreaks]{breakurl}

\usepackage{sectsty}
\allsectionsfont{\sffamily\mdseries\upshape} 

\usepackage[nottoc,notlof,notlot]{tocbibind} 
\usepackage[titles,subfigure]{tocloft} 

\title{\vspace{-2cm} An Empirical Comparison of Methods to Produce Business Statistics Using Non-Probability Data}
\author[1]{Lyndon Ang\thanks{Email:lyndon.ang@anu.edu.au}}
\author[2]{Robert Clark}
\author[3]{Bronwyn Loong}
\author[4]{Anders Holmberg}
\affil[1,2,3]{Australian National University, Canberra, Australia}
\affil[4]{Australian Bureau of Statistics, Canberra, Australia}

\date{September 2, 2024}

\begin{document}

\maketitle

\begin{abstract} 

There is a growing trend among statistical agencies to explore non-probability data sources for producing more timely and detailed statistics, while reducing costs and respondent burden.  Coverage and measurement error are two issues that may be present in such data.  The imperfections may be corrected using available information relating to the population of interest, such as a census or a reference probability sample.

In this paper, we compare a wide range of existing methods for producing population estimates using a non-probability dataset through a simulation study based on a realistic business population.  The study was conducted to examine the performance of the methods under different missingness and data quality assumptions.  The results confirm the ability of the methods examined to address selection bias.  When no measurement error is present in the non-probability dataset, a screening dual-frame approach for the probability sample tends to yield lower sample size and mean squared error results.  The presence of measurement error and/or nonignorable missingness increases mean squared errors for estimators that depend heavily on the non-probability data.  In this case, the best approach tends to be to fall back to a model-assisted estimator based on the probability sample.

\end{abstract}

Keywords - Estimation, Selection Bias, Big Data, Cut-off sampling, Dual frame

\newpage

\section{Introduction}

Probability survey samples have been, for the best part of the last century, the preferred data collection tool of statistical agencies for producing population estimates.  \citet{neyman_1934} laid the groundwork for design-based probability sampling theory, and the theory for estimating finite population quantities using probability survey samples is now well-established.  Compared with censuses, probability samples provide less costly, more timely, and more detailed information on populations of interest.  Methods have been developed to take advantage of additional information from other sources such as administrative datasets to improve the efficiency of estimates from probability samples (see, for example, \citealt{sarndal.etal_1992}).

In today's environment, the same reasons that drove the shift from censuses to probability samples are pushing statistical agencies to move away from probability samples to explore alternative data sources (\cite{beaumont_2020}).  Running a high-quality probability sample is costly, users have a desire for more timely statistics at greater levels of detail, and there is growing expectation for statistical agencies to reduce respondent burden.  For example, the Australian Bureau of Statistics (ABS) priorities for 2022-2023 include increasing the use of non-survey data sources to produce new, timely statistics and reduce burden on small and medium enterprises (\cite{australianbureauofstatistics_2022}).  Statistics Canada has launched a modernisation initiative with five pillars, one of which involves the use of new data sources, lower respondent burden, greater reliance on data integration and modelling and the reduced role of surveys (\cite{statisticscanada_2018}).

These new data sources may include big data or ``found data'' such as sensor data, satellite imagery and transactions data, and non-probability survey samples such as online web panels.  In this paper we apply the term non-probability dataset to any dataset where only part of the population is included, and the probability of inclusion into the dataset is unknown.

The potential benefits of non-probability data for producing statistics has been recognised, for example in  \citet{tam.clarke_2015}, \citet{rao.fuller_2017}, and \citet{tille.etal_2022}.  In particular, they promise to address the weaknesses of probability survey samples:

\begin{itemize}
	\addtolength\itemsep{-4mm}
	\item Data collection can be much cheaper since the data is already captured through an existing process (generally for some purpose independent of producing statistics)
	\item Outputs are more timely
	\item Utilising existing data reduces the need for agencies to ask people or businesses for the same or similar information
\end{itemize}

Unfortunately, non-probability data sources may suffer from quality issues.  Two types of error pervading these data are coverage error (also known as selection bias) and measurement error.  Coverage error occurs when we do not have a one-to-one correspondence between the population of interest and the population sampled.  It may include: undercoverage, where some units in the target population have been excluded from the sampling population (for example, new business startups not yet included on a business register at the time of a survey); overcoverage, where the sampling population includes units that are not in the target population (for example, inactive or closed businesses); and duplication, when some units in the sampling population can be selected more than once (for example, when a merger leads to the same business being included twice on a survey frame).

Our focus in this paper is on undercoverage error.  \citet{meng_2018} showed that undercoverage in a non-probability sample can have a significant impact on the quality of an estimate.  In fact, the error can get bigger the larger the dataset is, such that it may be preferable to use a small probability sample than a large dataset containing selection bias.

We will also refer to measurement error broadly as a misalignment between the true value of a target concept, and the value actually being captured by the sample.  The measurement error may thus arise due to issues such as differences in terms of how concepts are defined, recording or instrument errors, and so on.  For example, a business may report their revenue in whole dollars on a survey form which is asking for reporting to be done in thousands of dollars.

In recent years, methods have been developed to address the shortcomings of these new data sources, and in particular selection bias in the non-probability sample.  These methods enable the statistician to make direct use of the non-probability data to produce statistics.  Generally these methods require us to have additional information from the population that we can use.  The auxiliary information is typically used to correct for coverage error or to form models to impute for the missing units in the population, and could include known or estimated population totals, unit level information from a population register or frame, or from an independent probability survey sample (which we will refer to as a \textit{reference sample}) taken from the same population.

In this paper, we examine and compare through a realistic and wide-ranging simulation study the effectiveness of a cross-section of the estimation methods that have been developed for non-probability data, some in combination with probability sample data.  The comparison is undertaken in a business survey context.  Business surveys generally have a number of characteristics that will influence the choice and performance of an estimation approach; see \citet{hidiroglou.lavallee_2009} for a good overview of these characteristics.  For instance, there tends to be a larger quantity of auxiliary information available on a business population frame, including items that are well-correlated with the data items of interest, for example through administrative data from taxation agencies.  Business data items often also have skewed distributions, while their sample designs tend to be highly stratified, with large variations in the probabilities of selection.  In particular, the largest contributors tend to be included with a probability of 1.  Finally, there is often a reliable identifier (such as an official business number from the tax system) available, allowing different data sources to be linked.  These characteristics will influence the data available from any reference sample from a business population. 

We focus on the situation where the non-probability dataset makes up a large (approximately 50\%) portion of the population, an accompanying business probability sample from the population is available as auxiliary data, and the non-probability dataset can be linked to units on the probability sample and the population frame.  The study explores the effects of nonignorable non-response and measurement error in the non-probability sample on the performance of the various estimators, with the probability sample used to help address these issues.  To our knowledge, an empirical comparision of such a wide range of methods in a realistic business survey context has not been done before.  The results from the study provide some practical insights for statistical agencies looking to use these estimators to produce inference using their own large business non-probability datasets.

The rest of the paper is structured as follows.  The basic setup for this paper is outlined in Section \ref{setup}.  In Section \ref{np methods}, we provide a brief overview of the various estimation approaches that have been developed in recent years in the non-probability sample space.  Section \ref{sample designs} discusses a number of sample design frameworks that may be used to produce reference samples to address the shortcomings of the non-probability dataset.  In Section \ref{empirical} we provide a description of the empirical comparison study using simulated business data and discuss the results.  The simulation is grounded in data analysis of ABS Business Longitudinal Analysis Data Environment (BLADE) data and includes empirically-derived right-skewed distributions and realistic missingness scenarios in the non-probability data.  We conclude with some final thoughts in Section \ref{conclusion}.

\section{Basic Setup} \label{setup}

For each unit $i$ in a finite population $\mathcal{U}$ of size $N$, we have values $(\boldsymbol{x}_i,y_i)^T$ for a variable of interest $y$ and some additional auxiliary data items $\boldsymbol{x}$.  For the purpose of this paper we are interested in estimating the population total $Y=\sum_{i=1}^{N}y_i$, although we note it is also often of interest to estimate the mean $\bar{Y}=N^{-1}\sum_{i=1}^{N}y_i$.

Suppose we have a probability sample $A$ (the ``reference'' sample) of size $n_A$ drawn from the population.  $A$ may have been collected for a different purpose than we are interested in, and so may contain information about $\boldsymbol{x}$ only.  In other cases $A$ may include both $y$ and $\boldsymbol{x}$.  The data collected in $A$ are obtained without error.  Define $\pi^A_i=P(i \in A|\mathcal{U})$ as the inclusion probability for unit $i$ being in the probability sample, and $d^A_i=1/\pi^A_i$ is the design weight for $i \in A$.  The $\pi^A_i$ values are known from the sample design.

Denote a non-probability dataset by $B$.  Like the probability sample, $B$ may contain information on $\boldsymbol{x}$ only, for example if $B$ is an administrative dataset collecting general population information.  Alternatively $B$ may include both $\boldsymbol{x}$ and $y$, for example if it is a web panel survey collecting data on our variables of interest.  Let $\delta_i=I(i \in B)$ be an indicator variable for unit $i$ being included in the sample $B$.  The non-probability sample size is $N_B=\sum_{i=1}^{N}{\delta_i}$, $Y_B=\sum_{i=1}^{N} \delta_i y_i$ is the sum of the $y$ values in the non-probability dataset, and $\boldsymbol{X}_B=\sum_{i=1}^{N} \delta_i\boldsymbol{x}_i$ is the sum of the $\boldsymbol{x}$ variables in the non-probability dataset.  In contrast to $\pi^A_i$ in the reference probability sample, the inclusion probabilities $\pi^B_i=P(\delta_i=1|\mathcal{U})$ are unknown and need to be estimated.  Further define $C=\mathcal{U} \setminus B$, the units in the population not included in $B$.  There may or may not be overlap between the two samples $A$ and $B$.  Figure \ref{fig:population domains} depicts these domains within $\mathcal{U}$.

\tikzset{filled/.style={fill=circle area, draw=circle edge, thick},
    outline/.style={draw=circle edge, thick}}

\setlength{\parskip}{5mm}

\begin{figure}[!h]
	\centering

	\colorlet{circle edge}{black!50}
	\colorlet{circle area}{black!20}

	\tikzset{
  		filled/.style={fill=none, draw=circle edge, thick},
  		outline/.style={draw=circle edge, thick}
	}

	\begin{tikzpicture}
		\draw[filled] (0,0) circle[radius=1.2cm] node {$A \setminus B$}
					  (0:2cm) circle[radius=1.2cm] node {$B \setminus A$};
		\node[anchor=north] at (current bounding box.south) {$A \cap B$};
		\draw[thick, ->] (1,-1.3) -- (1,-0.2);
		
		\node [draw,fit=(current bounding box),inner sep=6mm] (frame) {}; 
		\node [below left] at (frame.north west) {$\mathcal{U}$};
		\node[above right] at (frame.south west) {$C \setminus A$};

	\end{tikzpicture}
	\caption{Domains Within the Population $\mathcal{U}$}
	\label{fig:population domains}
\end{figure}
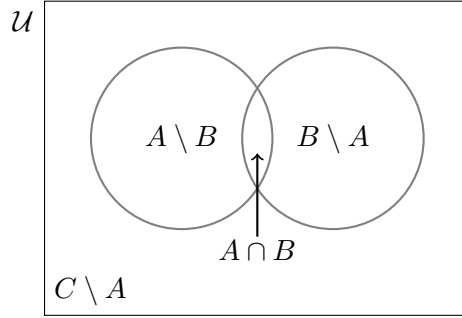

Some assumptions are often made regarding the selection mechanism into $B$, in order to facilitate inferences using those datasets.  Three assumptions are generally adopted by the methods in this paper; see, for example \citet{chen.etal_2020} and \citet{yang.etal_2021}.  These assumptions include:

\begin{enumerate}
    \addtolength\itemsep{-4mm}
	\item[]
    \begin{enumerate}
        \addtolength\itemsep{-4mm}
        \item \textit{Ignorability}: Conditional on the set of covariates $\boldsymbol{x}_i$, $\delta_i$ and $y_i$ are independent.
        \item \textit{Positivity} or \textit{Common Support}:  Conditional on $\boldsymbol{x}_i$, $P(\delta_i=1|\boldsymbol{x}_i) > 0, \forall i \in \mathcal{U}$.
        \item \textit{Independence}: Conditional on $\boldsymbol{x}_i$ and $\boldsymbol{x}_j$, $\delta_i \perp \delta_j, i \neq j, \forall i,j \in \mathcal{U}$.
    \end{enumerate}
\end{enumerate}

Ignorability implies that $P(\delta_i=1|\boldsymbol{x},y)=P(\delta_i=1|\boldsymbol{x})$.  In other words, selection into $B$ is ignorable conditional on the covariates $\boldsymbol{x}$.  This assumption is similar to the Missing-At-Random (MAR) scenario of \citet{rubin_1976}.  \citet{andridge.etal_2019} and \citet{boonstra.etal_2021} refer to this type of selection process as Selection At Random (SAR).  When selection into $B$ is influenced by $y$, then we instead have Selection Not At Random (SNAR), akin to the Missing-Not-At-Random (MNAR)/Not-Missing-At-Random (NMAR) scenario.  In our paper we adopt the SAR and SNAR terminology to describe the type of missignness in $B$.  The simulation study explores how estimators perform under the SAR versus SNAR situations.

In addition to the above, the following will be assumed in some of the methods discussed in this paper:

\begin{enumerate}
    \addtolength\itemsep{-4mm}
	\item[]
    \begin{enumerate}
        \addtolength\itemsep{-4mm}
        \setcounter{enumii}{3}
        \item It is possible to accurately link records in $B$ to the corresponding unit in the population frame and $A$.
        \item There is full response in $A$.
        \item The full set of auxiliary information $\boldsymbol{x}$ is available in both $A$ and $B$, without error.
    \end{enumerate}
\end{enumerate}

Assumption A4 may be satisfied if there exists a common unit identifier available on $B$ and the population frame, or if it is possible to accurately link records in $B$ to the population frame and/or $A$ using a common set of linking variables.  In the business population setting, businesses may have a unique business number for administrative purposes such as filing taxation returns, and this identifier can be available on both the business frame and external data sources.

Assumption A5 will tend not to reflect reality, especially in this age of declining response rates (\citealt{beaumont_2020}).  The full response assumption in $A$ is made to simplify the simulation and discussion of its results, but we note that in general this assumption may be relaxed by assuming response probabilities in $A$ can be estimated using $\boldsymbol{x}$ or a subset thereof.  The relaxation implies that any non-response in $A$ will be SAR and addressable through a non-response weight adjustment or imputation process.  With sufficient resources dedicated to careful non-response follow-up, we should be able to ensure response rates are relatively high and render any remaining non-response in the probability sample as ignorable.

Assumption A6 implies that we have access to all auxiliary variables relevant to Assumptions A1, A2 and A3 on the datasets $A$ and $B$, and these variables can be fed in without error into any models we subsequently develop.  On the other hand, we allow $y$ to be collected with measurement error on the non-probability dataset.  Instead of observing $y_i$ we observe $y^*_i$, and the two are assumed to be related through the relationship (see \citealt{kim.tam_2021})

\begin{equation} \label{eqn: me defn}
    y^*_i = \beta_0 + \beta_1y_i + e_i
\end{equation}

where $(\beta_0,\beta_1)$ is unknown and $e_i \sim (0,\sigma^2)$.  In the case that $(\beta_0,\beta_1)=(0,1)$, we have no measurement error in $y$.  In our simulation study, we are particularly interested in the performance of estimators with versus without measurement error.

\section{Inference methods for non-probability datasets} \label{np methods}

In this section we provide a brief overview of the literature on methods developed for making inference using non-probability datasets.  A number of very good review papers also exist in this space that provide more detailed discussions of the various methods.  The interested reader is pointed to \citet{lohr.raghunathan_2017}, \citet{zhang_2019}, \citet{tam.holmberg_2020}, \citet{yang.kim_2020}, \citet{rao_2021} and \citet{wu_2022a}.  More recently, \citet{salvatore_2023} analysed a large number of documents on this topic using text mining and bibliometric techniques to identify current research trends.

The method of inference used depends on the data structures we have for the non-probability and reference samples.  There are three general approaches that will be discussed in this paper: weighting-based approaches, imputation-based approaches, or a combination of the two (so-called doubly-robust methods).

The data available to us in $A$ and $B$ will influence what estimation methods will be most suitable to use.  Table \ref{tab:data structures}, inspired from \citet{tam.holmberg_2020}, describes four (Type I - Type IV) potential data structures relating to the non-probability sample and an accompanying probability sample.  For completeness, we also include a fifth data structure - Type 0 - which reflects the more standard survey situation when data is available from a probability sample and accompanying population frame, but not the non-probability sample.  We will occasionally refer to these data types in our discussion.

\begin{table}[h!]
	\begin{center}
		\caption{Data structures}
		\label{tab:data structures}
		\begin{tabular} {c c c c} \toprule
			\multicolumn{2}{c}{\textbf{Data type}} & \multicolumn{1}{c}{\textbf{Response}} & \multicolumn{1}{c}{\textbf{Auxiliary}} \\
			\multicolumn{2}{c}{} & \multicolumn{1}{c}{\textbf{Variables $y$}} & \multicolumn{1}{c}{\textbf{Variables $\boldsymbol{x}$}} \\
			\toprule
			\multirow{2}{*}{Type 0} & \multicolumn{1}{l}{Frame} & $\times$ & \checkmark  \\
									& \multicolumn{1}{l}{P Data $(A)$} & \checkmark & \checkmark  \\\cmidrule{1-4}
			Type I & NP Data $(B)$ & \checkmark & \checkmark \\\cmidrule{1-4}
			\multirow{2}{*}{Type II} & \multicolumn{1}{l}{NP Data $(B)$} & \checkmark & \checkmark \\
									& \multicolumn{1}{l}{P Data $(A)$} & $\times$ & \checkmark \\\cmidrule{1-4}
			\multirow{2}{*}{Type III} & \multicolumn{1}{l}{NP Data $(B)$} & \checkmark & \checkmark \\
									& \multicolumn{1}{l}{P Data $(A)$} & \checkmark & \checkmark \\\cmidrule{1-4}					   				
			\multirow{2}{*}{Type IV} & \multicolumn{1}{l}{NP Data $(B)$} & $\times$ & \checkmark  \\
									& \multicolumn{1}{l}{P Data $(A)$} & \checkmark & \checkmark \\			
			\bottomrule
		\end{tabular}
	\end{center}
\end{table}

\subsection{Weighting approaches} \label{weighting approaches}

Weighting approaches create a weight associated with each record in the reference sample or the non-probability sample, and these are used to form estimates for a target population quantity using an Inverse Probability Weighting (IPW) approach.  Define $w^A_i$ as the survey weight associated with a record $i$ in $A$.  $w^A_i$ may be equal to $d^A_i=1/\pi^A_i$ defined in Section \ref{setup}, or $d^A_i$ adjusted via a non-response and/or calibration process.  Population totals may be estimated by $\hat{Y}=\sum_{i \in A} w^A_iy_i$.

Let $\hat{\pi}^B_k$ be an estimate of the propensity of selection into $B$ for unit $k$ in $B$.  Then $w^B_k=1/\hat{\pi}^B_k$ is the corresponding weight associated with record $k$, and we may form estimates of population totals using $\hat{Y}=\sum_{k \in B} w^B_ky_k$.  \citet{wu_2022a} notes that a better estimator to use is the H\'ajek estimator

\begin{equation} \label{eqn:Hajek IPW estimator}
	\hat{Y}_{H\acute{a}jek}=\frac{N}{\hat{N}_B} \sum_{k \in B} w^B_ky_k
\end{equation}

where $\hat{N}_B=\sum_{k \in B} w^B_k$.

When a Type III data structure exists, and assuming that we can determine the value of $\delta_i$ for the units in $A$ (for example, by linking the sample $A$ to $B$), \citet{kim.tam_2021} proposed calibrating the design weights in $A$ to the quantities $\sum{_{i=1}^{N}(1,\delta_i,\delta_iy_i)=(N,N_B,Y_B)}$.  The calibration process involves finding new weights $w^A_i$ for $i \in A$ that minimise a chosen distance metric subject to the calibration constraints $\sum_{i \in A}w^A_i\boldsymbol{x}_i=\sum_{i \in \mathcal{U}}\boldsymbol{x}_i = \boldsymbol{X}$ (\citealt{deville.sarndal_1992}), where $\boldsymbol{X}$ is the vector of population totals for the set of auxiliary variables.  In particular, the generalised regression estimator is a special case of the calibration estimator when the Chi-Square distance metric $\sum_{i \in A}(w^A_i-d^A_i)^2/2q_id^A_i$ is used, where $q_i$ is a tuning parameter.  The set of calibration constraints can be expanded to include auxiliary variables (for example to address non-response) in $A$, and the method can also cater for measurement error in $y$ on either $A$ or $B$.

Rather than making a weight adjustment for the probability sample, we can instead produce a weighted non-probability sample.  This seems attractive in particular when $N_B$ is large in comparison with $n_A$.  The main purpose of the reference sample $A$ is now to aid in estimating the propensities of selection $\pi^B_k=f(\boldsymbol{x}_k,\boldsymbol{\phi})$ for $k \in B$, where $f$ is a chosen parametric form such as the inverse logit function, and $\boldsymbol{\phi}$ are unknown parameters requiring estimation.

Under Assumption A1 and a Type I data structure, selection bias may be reduced by applying a weight calibration process to $B$ which satisfies $\sum_{k \in B}w^B_k\boldsymbol{x}_k=\boldsymbol{X}$ (see, for example, \citealt{haziza.etal_2010}).  The initial weights feeding into the calibration may be given by $d^B_k = N/N_B$ or $d^B_k = 1$ (\cite{rueda.etal_2022}, \cite{golini.righi_2024}).  If $\boldsymbol{X}$ is not available but a Type III data structure exists, a pseudo-calibration estimator may be applied (\cite{golini.righi_2024}; see also \cite{righi.etal_2019}) using estimated totals from the reference probability sample to feed in to the calibration equation

\begin{equation} \label{eqn:PC estimator}
    \sum_{k \in B}w^B_k\boldsymbol{x}_k=\sum_{i \in A}d^A_i\boldsymbol{x}_i
\end{equation}

to produce final weights $w^B_k$ for the units in $B$.

Direct estimation of propensity scores $\hat{\pi}^B_k$ may be accomplished in a variety of ways, and methods generally assume a Type II data structure.  \citet{chen.etal_2020} derive a pseudo log-likelihood function with two terms, one involving $B$ and the other involving $A$, and this function can be maximised through iterative methods such as the Newton-Raphson procedure.  \cite{burakauskaite.ciginas_2023} detail a variation of this method which is applicable when $\boldsymbol{x}_i$ are available for $i \in \mathcal{U}$.  \citet{kim.wang_2019} assume that it is possible to determine $\delta_i$ for units in $A$, and use this to estimate propensity scores based on the probability sample.  

\citet{elliott.valliant_2017} estimate $\hat{\pi}^B_k$ for $k \in B$ by first pooling together the non-probability data and reference probability data.  For unit $i$ in the pooled dataset $A \cup B$, they define $Z_i=1$ if unit $i$ belongs to $B$, and $Z_i=0$ if unit $i$ belongs to $A$.  Estimated conditional probabilities $\hat{P}(Z_i=0|\boldsymbol{x}_i)$ and $\hat{P}(Z_i=1|\boldsymbol{x}_i)$ are estimated through a modelling process, and then combined to produce an estimated propensity of selection for units in $B$.  This approach does not require knowledge about $\delta_i$ for units in $A$.  On the other hand, the method assumes small sampling fractions and no overlap in the units captured within the two samples - an assumption that is not likely to hold if $B$ is large.  \citet{liu.etal_2022} addressed the overlap issue by using only the non-overlapping units in the pooled sample to fit a model to estimate certain probabilities to feed into the creation of pseudo-weights.  Their method requires knowing which units comprise $A \cap B$ (so that these can be removed from the pooled sample), and is not appropriate when the non-overlapping portion of the two samples is very small, or if one sample is a subset of the other.

\citet{wang.etal_2021} proposed an Adjusted Logistic Propensity (ALP) weighting method which also pools the non-probability and probability datasets together, but does not require the assumption of non-overlapping samples.  However, in their method the final estimated propensity may sometimes be greater than 1 if $B$ is large in size.  \citet{savitsky.etal_2023a} noted that the methods of \citet{wang.etal_2021} and \citet{chen.etal_2020} are sub-optimal as they rely on pseudo-likelihoods for estimating propensities.  Instead, \citet{savitsky.etal_2023a} constructed a likelihood defined directly on the observed pooled sample to estimate propensities.  Their method does not require the samples $A$ and $B$ to be disjoint, allowing an unknown amount of overlap between the samples to be present.  A hierarchical Bayesian approach was utilised to enable computation of all required probabilities simultaneously.  In a simulation setting under moderate sample sizes, the method was found to produce more accurate estimates of inclusion probabilities for $B$ compared to the pseudo-likelihood approaches used in \citet{wang.etal_2021} and \citet{chen.etal_2020}.

If Assumption A1 does not hold and the missingness in $B$ is SNAR, then $\pi^B_k=f(\boldsymbol{x}_k, y_k, \boldsymbol{\phi})$ for $k \in B$ and we need to include the variable of interest $y$ in the model to estimate $\hat{\pi}^B_k$.  \citet{marella_2023} and \citet{kim.morikawa_2023} employed the sample empirical likelihood to estimate $\hat{\pi}^B_k$ under a SNAR selection mechanism for $B$.  If known population means of auxiliary variables are available under a Type I data structure, they can be included as calibration constraints in the maximisation process for the empirical likelihood to help address selection bias.  \citet{marella_2023} noted that under a Type II data structure, sample estimates for these auxiliary variables may be used as the calibration constraints.

Machine learning techniques have also been explored to estimate the propensity scores.  \citet{ferri-garcia.rueda_2020} provide a comparison of different machine learning approaches using a pooled dataset $A \cup B$ which assumes no overlapping sample, and under a Type II data structure.  \citet{castro-martin.etal_2022} proposed including the weights derived from a propensity score estimation process, $\hat{w}_k^B=1/\hat{\pi}^B_k, k \in B$, as part of a subsequent machine learning model training process (for example, linear regression) to predict $\hat{y}_i, i \in A$ under a Type II data structure.  The imputed values obtained from the trained model may then be used to form estimates based on $A$:

\begin{equation} \label{eqn:MI1}
	\hat{Y}_{MI1}=\sum_{i \in A} w^A_i\hat{y_i}
\end{equation}

More accurate estimated propensity scores may be achieved by incorporating known information about the auxiliary variables in the propensity score estimation for $B$.  \citet{zhu.etal_2023} assume a latent Gaussian copula model for the joint distribution of the auxiliary variables.  The model is fitted using data in $A$, and a pseudo-population is simulated using the fitted model.  The pseudo-population is used to estimate the marginal inclusion probabilities $P(\delta_k=1), k \in B$, and these are then used to estimate propensity scores $\hat{\pi}^B_k$.

\subsection{Imputation-based approaches} \label{imputation approaches}

Imputation-based approaches assume that we can form a reliable estimate for $y_i$ using the available auxiliary information $\boldsymbol{x}_i$.  This approach will generally be used when $y$ is missing from $A$ (Type II data structure) or $B$ (Type IV data structure).  When a Type II data structure exists, the model for $y$ is formed using the data in the non-probability sample and then applied to impute $\hat{y}$ for all the units in the reference sample, referred to as mass imputation (\citealt{chipperfield.etal_2012}).

Under a Type II structure, an alternative imputation estimator to (\ref{eqn:MI1}) is (see \citealt{wu_2022a})

\begin{equation} \label{eqn:MI2}
	\hat{Y}_{MI2} = \sum_{k \in B}y_k + \Bigl(\sum_{i \in A}w^A_i\hat{y}_i - \sum_{k \in B}\hat{y}_k\Bigr)
\end{equation}

where $w^A_i$ may be design, non-response adjusted, or calibrated weights for $i \in A$, and $\hat{y}_i$ are predictions from a model, which may range from a linear model, a semi-parametric model such as a generalised additive model or a kernel regression (\cite{chen.etal_2022a}), to non-parametric methods such as regression trees and random forests (\cite{golini.righi_2024}).  The estimator (\ref{eqn:MI2}) may be interpreted as the sum of the true values from $B$ and an estimated contribution for $C=\mathcal{U} \setminus B$ based on the modelled $\hat{y}$ values.

\citet{rivers_2007} proposed a sample matching approach using nearest neighbour imputation to mass impute the missing $y_i$ values for $i \in A$ under a Type II data structure.  The non-probability dataset is treated as the donor population.  A distance measure indicating the similarity of units in $A$ and $B$ is calculated using the available covariates $\boldsymbol{x}$, and the closest match is chosen to supply $\hat{y}_i$.  \citet{yang.etal_2021} extended the nearest neighbour imputation method to $k$ nearest neighbours, again assuming a Type II data structure, where data from the $k$ nearest neighbours are combined to produce a mean value $\hat{y}_i = k^{-1}\sum_{i=1}^{k}y_i$ which is used as the impute.  When $\delta_i$ can be identified for each unit in $A$, the authors showed that the efficiency of the nearest neighbour imputation can be improved by also calibrating the weights in the probability sample to match known quantities in $B$, for example:

\begin{equation}
    \sum_{i \in A}w^A_i(\delta_i,1 - \delta_i,\delta_i\boldsymbol{x}_i,\delta_i \hat{y}_i) = (N_B, N - N_B, \boldsymbol{X}_B, Y_B) \nonumber
\end{equation}

where $w^A_i$ are the final calibrated weights and $N$, $N_B$, $\boldsymbol{X}_B$ and $Y_B$ are defined in Section \ref{setup}.

When a Type IV data structure exists, the reference sample may be used to estimate the parameters of the model for $y_i|\boldsymbol{x}_i$.  The fitted model is then used to provide predictions for $y_k, k \in B$.  In \citet{righi.etal_2019} and \citet{golini.righi_2024}, a combined imputation and weighting approach is proposed whereby predicted values $\hat{y}_k$ for the variable of interest are produced from a modelling process, and are used in a pseudo-calibration estimator $\hat{Y}_{PC}=\sum_{k \in B}w^B_k\hat{y}_k$, where $w^B_k$ is the solution to the calibration equation (\ref{eqn:PC estimator}).  The authors note that their approach can also be used in a Type III data structure when $y$ is observed with error in $B$ (that is, $B$ collects $y^*_k$ instead of $y_k$) and a measurement error model such as (\ref{eqn: me defn}) is fitted to correct the error.

In the case where the modelling process fails to capture the true relationship between predictors and the variable of interest, $\hat{Y}_{PC}$ will be biased.  \citet{righi.etal_2019} and \citet{golini.righi_2024} proposed amending $\hat{Y}_{PC}$ by including a bias correction term in the estimator.  This leads to the difference estimator (see also \citealt{breidt.opsomer_2017})

\begin{equation} \label{eqn:diff1_PC}
    \hat{Y}_{DPC} = \sum_{k \in B}w^B_k\hat{y}_k + \sum_{i \in A}d^A_i(y_i - \hat{y}_i)
\end{equation}

\citet{medous.etal_2023a} extended the calibration approach of \citet{kim.tam_2021} from a Type III to a Type IV structure, proposing so-called QR predictors (see \citet{wright_1983}) to produce an estimated $\hat{Y}_{B}$.  $\hat{Y}_B$ may then be combined with an estimated contribution from $C$ (via $A$) to produce an improved population estimate.

\subsection{Doubly Robust estimation}

Many of the approaches outlined in the previous sections depend on an accurate working model.  To protect against model misspecification, some authors have suggested doubly robust estimators constructed using both a propensity score model for $B$ and a model for $y|\boldsymbol{x}$.  The setup only requires one of the two models to be correctly specified in order to be unbiased.

Assuming a Type II data structure, \citet{chen.etal_2020} proposed two doubly robust estimators for $\bar{Y}$, with the second estimator preferred:

\begin{equation} \label{eqn:DR1}
	\hat{\bar{Y}}_{DR1} = \frac{1}{N} \sum_{k \in B} \frac{y_k-\hat{y}_k}{\hat{\pi}^B_k} + \frac{1}{N} \sum_{i \in A} \frac{\hat{y}_i}{\pi^A_i}
\end{equation}

and

\begin{equation} \label{eqn:DR2}
	\hat{\bar{Y}}_{DR2} = \frac{1}{\hat{N}^B} \sum_{k \in B} \frac{y_k-\hat{y}_k}{\hat{\pi}^B_k} + \frac{1}{\hat{N}^A} \sum_{i \in A} \frac{\hat{y}_i}{\pi^A_i}
\end{equation}

where $\hat{y}_i$ and $\hat{y}_k$ are the estimated values of $y_i, i \in A$ and $y_k, k \in B$ based on a fitted imputation model using $(y_k,\boldsymbol{x}_k)$ in B, $\hat{\pi}^B_k$ is the estimated probability of inclusion for $k \in B$ using a method from Section \ref{weighting approaches}, $\hat{N}_A=\sum_{i \in A}w^A_i$ and $\hat{N}_B=\sum_{k \in B}1/\hat{\pi}^B_k$.

The doubly robust approach can be extended to a multiply robust scenario.  Instead of employing a single propensity score model to estimate $\pi^B_k,k \in B$, and a single imputation model, \citet{chen.haziza_2022} suggested the use of $m$ propensity score models and $m$ imputation models.  Each model may be based on different sets of explanatory variables.  The results from the $m$ models are ``compressed'' or summarised to form an overall estimated value for $\hat{y}_k$ or $\hat{\pi}^B_k$.  The estimator is consistent as long as one of the imputation or propensity score models is correctly specified.  \citet{kim.morikawa_2023} also suggest the use of multiple propensity score models and multiple constraints for bias calibration in their empirical likelihood approach.

\section{Alternative sampling frameworks for the reference sample} \label{sample designs}

In the approaches described in Section \ref{np methods}, the reference probability samples $A$ are assumed to be a given.  In practice, the reference sample design may be approached in a number of ways based on the nature of the available non-probability data.  In this section, we briefly discuss two sampling frameworks for the reference sample as alternatives to a traditional design-based sample from the full population.

\subsection{Multiple frame approach}

The original theory for multiple-frame surveys was developed by \citeauthor{hartley_1962} (\citeyear{hartley_1962,hartley_1974}), and has been built upon in recent years; see, for example, \citet{lohr.rao_2006} and \citet{lohr_2011}.  In the context of non-probability data, one can consider $B$ to be a full census from an incomplete population ``frame'' (\cite{lohr_2021}), and may or may not measure $y$, instead measuring covariates $\boldsymbol{x}$ that can be used to predict $y$.  \citet{lohr_2021} and \citet{medous.etal_2023a} note that the data integration estimators developed in \citet{kim.tam_2021} can be considered within a multiple frame context.

Assume that we can identify $C = \mathcal{U} \setminus B$, for example through a common unit identifier (in general, the multiple frame approach assumes that we can determine which population segment(s) each unit belongs to).  Then we can employ a \textit{screening} dual frame sample design and select a probability sample $A$ from $C$ only.  A consistent estimator of $Y$ is then simply

\begin{equation} \label{eqn:Split Popn}
	\hat{Y}_{screening} = Y_B + \hat{Y}_C
\end{equation}

where $\hat{Y}_C$ may be estimated using a Horvitz-Thompson or H\'ajek estimator.  Calibration benchmarks for the subpopulation $C$ may be used if available to improve the efficiency of $\hat{Y}_C$.  \citet{zhang_2019} refers to (\ref{eqn:Split Popn}) as the \textit{split-population approach} to inference, where $B$ and $C$ constitute the two ``populations''.  More generally, \citet{zhang_2019} noted that one can produce a composite estimator for the population mean based on the mean of $B$ and the mean of $C$ estimated from a reference sample.

A number of assumptions are generally made when making inferences under the multiple-frame framework.  Within a data integration context, some of these assumptions may be less likely to hold.  One potential issue is that the variables captured on the non-probability source may not match exactly the variables of interest captured in the probability sample.  If a ``screening'' dual frame design has been used, the lack of overlapping sample makes it more difficult to assess and address any measurement error in the non-probability dataset.

\subsection{Cut-off sampling}

In cut-off sampling, a certain fraction of the population are deliberately excluded from the survey frame.  A probability sample is then taken from the remainder of the population.  This practice tends to be used in business surveys when the variable of interest is highly skewed; \citet{yorgason.etal_2011} and \citet{elisson.elvers_2001} provide some examples.  Generally, the smallest businesses are placed into a single ``take-none'' stratum with zero probability of selection.  The contribution of these businesses to the variable of interest is assumed to be negligible compared with the remaining part of the population, so there is a saving in terms of reduced respondent burden and cost without causing significant bias (\citealt{elisson.elvers_2001}).  The rest of the population may be further divided into a ``take-all'' (completely enumerated) and a ``take-some'' (sampled) stratum.  Denote the population in the take-none, take-some and take-all strata as $\mathcal{U}_E$, $\mathcal{U}_S$, and $\mathcal{U}_{CE}$ respectively, while $Y_E$, $Y_S$ and $Y_{CE}$ are the totals for $y$ in those strata.

The non-probability dataset would often be useful for estimating the part of the population $\mathcal{U}_E$ intentionally not covered by the cut-off sample A. For example, if $y$ is measured in both $A$ and $B$, then we might estimate $Y_E$ as

\begin{equation}
	\hat{Y}_E = \frac{N_E}{N_{E,B}} \sum_{i \in \mathcal{U}_E \cap B} y_i \nonumber
\end{equation}

where $N_E$ is the number of units in the population in the take-none stratum and $N_{E,B}$ is the number of units in $B$ in the take-none stratum.

Note that this estimator will only be approximately unbiased if $\frac{Y_E}{Y_{CE}+Y_S} \approx \frac{Y_{B,E}}{Y_{B,CE}+Y_{B,S}}$. This approximation may be very imperfect in practice, but it may still be adequate when $Y_E$ is small, as is usually the case by design. If we have available auxiliary variables $\boldsymbol{x}$ for the whole population, then we may use this to model the propensities $\pi_k^B$, and thus form an IPW estimate for the take-none stratum by applying (\ref{eqn:Hajek IPW estimator}) to $B \cap \mathcal{U}_E$.  The audit sampling ideas of \citeauthor{zhang_2019} (\citeyear{zhang_2019, zhang_2021a, zhang_2023a}) may come in useful here to inform appropriate action for the cut-off population, with the idea that the audit sample only has to be taken intermittently.

\section{Empirical comparison of estimation approaches} \label{empirical}

A simulation study was conducted to compare different estimation approaches for non-probability samples within a business survey context.  We selected a cross-section of approaches that would be straightforward to implement for a statistical agency.  The aim of the exercise was to examine their performance under four scenarios: SAR vs SNAR missingness in the big dataset, and with vs without measurement error in $y$ on the big dataset.  In our simulation the non-probability dataset includes a large fraction (about 50\%) of the population, so may be considered a ``big dataset''.

In addition to Assumptions A4 to A6 in Section \ref{setup}, the simulation study assumes that the population frame includes some auxiliary information, including frame employment and industry class, and these variables are available to use during the estimation process.  We can reasonably expect this to hold in a business survey context.  For example, some business information is often be available from administrative sources (like business tax data) to attach to the population frame.

The simulation consisted of the following steps: (1) generating a number of data items for a finite population with distributional properties similar to some items on a real business survey dataset, (2) drawing a random subsample from the population to be the ``big dataset'', (3) drawing a reference probability sample, and (4) applying some of the methods described in Section \ref{np methods} to produce estimates.

For each of the sample designs considered, $R=2,000$ repeated samples $A$ and $B$ were drawn and combined in some way to produce estimates.  The Monte Carlo Relative Bias (RB) and Relative Root Mean Squared Error (RRMSE) of the estimators were then calculated as

\[ \text{RB} = \frac{1}{R}\sum_{r=1}^{R}\frac{\hat{Y}_r - Y}{Y} \]
\[\text{RRMSE}=\frac{\sqrt{\frac{1}{R}\sum_{r=1}^{R}(\hat{Y}_r-Y)^2}}{Y} \]

where $\hat{Y}_r$ represents the estimate of $Y$ from the $r$'th repeated sample, and $Y$ is the true population total for $y$.

\subsection{Generating the population and big dataset}

The simulated population has $N=900,000$ business records, consisting of two categorical domain variables (State and Industry), a frame size variable (Frame Employment), and three survey variables (Reported Employment, Total Weekly Wages/Salaries, and Overtime Pay).  It resembles the real-world population of employing businesses in Australia in terms of the distribution of businesses across size categories, industry divisions and state.  A combination of published survey outputs alongside employee tax data and survey microdata sourced from the ABS DataLab environment was used to help generate the survey variables of interest.  The Supplemental Data provides a detailed description of the process used to create the population and the data items with and without measurement error.  The synthetic population is available at this link: \url{https://zenodo.org/records/11095755}.

The selection mechanism for inclusion in the big dataset is given by $\delta_k \sim \text{Bernoulli}(\pi^B_k)$.  A two-stage process was used to generate the final values of $\pi^B_k$ for the population.   At the first stage, an initial probability $\pi^B_{1k}$ was produced, where

\[ \pi^B_{1k} = \frac{\text{exp}(\phi_0 + \phi_1x_{k} + \phi_2y_{k})}{1 + \text{exp}(\phi_0 + \phi_1x_{k} + \phi_2y_{k})} \]

Two types of big dataset were produced, one following a SAR process, and one following a SNAR process.  $(\phi_0,\phi_1,\phi_2)=(0.09,0.009,0)$ for the SAR dataset, and $(\phi_0,\phi_1,\phi_2)=(0.85,0.009,-0.1)$ for the SNAR dataset.  In the calculation of $\pi^B_{1k}$ for both datasets, the $x$ variable used was Frame Employment, and the $y$ variable used was the natural logarithm of total weekly earnings.

At the second stage, the $\pi^B_{1k}$ values were adjusted downwards by a pre-specified factor in some industries, to simulate a reduced likelihood of being present on the big dataset for those industries.  The resulting probabilities were our final $\pi^B_k$ values.  In our non-response models, units with smaller frame employment have a lower chance of being included in the big dataset $B$.

For each sample draw of the simulation, a big dataset sample was drawn using Poisson sampling and the final $\pi^B_k$ probabilities.

\subsection{Probability sample designs} \label{probability sample designs}

The population was assigned to strata based on State, Industry Division and Frame Employment for each business.  Size stratum categories were: 0-4 employees, 5-19 employees, 20-299, and 300+.

Three sample design scenarios were examined:

\begin{itemize}
	\addtolength\itemsep{-4mm}
	\item Single-frame - An optimal allocation of sample to strata using the full population frame, $\mathcal{U}$
	\item Dual-frame - An optimal allocation of sample to strata using $C = \mathcal{U} \setminus B$ as the probability sampling frame (with $B$ comprising the other frame)
	\item Cut-off - An optimal allocation of sample to strata where the sampling frame is the population excluding units in the smallest (0-4 employees) size class
\end{itemize}

The Bethel-Chromy algorithm (\citealt{bethel_1989}, \citealt{chromy_1987}) was applied to produce optimal allocations for each of the three scenarios, treating the relevant sample frame as the population of interest.  For example, in the dual-frame scenario the algorithm was applied to achieve accuracy targets based on the units in $C$.  The sample designs were produced to meet the following accuracy constraints for the total earnings data item on the relevant sample frame:

\begin{itemize}
	\addtolength\itemsep{-4mm}
	\item Relative Standard Error (RSE) of 1.5\% at the National level
	\item RSE of 5\% for each Industry Division
	\item RSE of 5\% for each State
\end{itemize}

A minimum sample size of 6 was applied for each sampled stratum.  The 300+ size strata were designated to be completely enumerated strata with a sampling fraction of 1.

\subsection{Estimators examined}

Table \ref{tab:estimator desc} provides a description of the estimators compared for the single-frame sample design.

\begin{table}[!th]
	\begin{center}
		\caption{Summary of estimators compared - single frame sample}
		\label{tab:estimator desc}
		\begin{tabular} {p{2.5cm} p{1.7cm} p{9cm}} \toprule
			Estimator & Data Scenario & Description \\
			\toprule
				GREG  & Type 0 & Generalised Regression estimator with frame employment as the $x$ variable    \\   
				RDI  & Type III & The Regression Data Integration estimator of \citet{kim.tam_2021}, where the probability sample is calibrated to big data totals $(N,N_B,Y_{earnings})$.  In the with-measurement error scenario, the probability sample is calibrated to the measurement error versions of the data item    \\
				QR MA & Type IV & The model-assisted QR estimator described in \citet{medous.etal_2023a}.  Frame employment is used as the explanatory variable \\
				KW & Type II & Estimation of $\hat{\pi}^B_k, k \in B$ by employing the method of \citet{kim.wang_2019}.  $\delta_i, i \in A$ is obtained by linking units in $A$ and $B$.  Frame employment and industry division are the explanatory variables in the model for $\hat{\pi}^B_k$.  Estimates are produced using (\ref{eqn:Hajek IPW estimator}).  In the with-measurement error scenario, the measurement error versions of the data items are used in estimation    \\
				KW-Cal & Type II & A two-step weighting process starting with creation of KW weights (see above).  These weights are then calibrated to population size $N$ and frame employment total.  The with-measurement error scenario uses the measurement error versions of the data items on the big dataset \\
				KW-Earn & Type II & The KW estimator with the addition of the natural logarithm of Earnings as an explanatory variable in the model for $\hat{\pi}^B_k$ \\
				ALP & Type II &  The Adjusted Logistic Propensity method of \citet{wang.etal_2021}, where estimation of $\hat{\pi}^B_k, k \in B$ involves pooling data in $A$ and $B$ together and fitting a weighted logistic model on the pooled data \\
				Wgt\_Reg\_MI & Type II & A mass imputation for $y$ in $A$ using weighted regression modelling.  We assume in this case that $y$ variables are not available in $A$.  We fit weighted models for each $y$ data item based on the big data, where the weights are the estimated KW weights.  The models are applied to the units in the probability sample to produce $\hat{y}_i, i \in A$.  (\ref{eqn:MI1}) is used to produce estimates of total \\
				DR\_wgt & Type II & The doubly-robust estimator (\ref{eqn:DR2}) which combines the KW and Wgt\_Reg\_MI approaches \\
				HD\_MI & Type II & Hot deck mass imputation method to impute $y_i, i \in A$, where industry and size groups are used to form the classes that the hot deck imputation will be performed within \\
			\bottomrule
		\end{tabular}
	\end{center}
\end{table}

The dual-frame probability sample design is a \textit{screening} dual-frame design.  This allows us to combine the estimate from the probability sample with the big data total to form an estimate for $\mathcal{U}$.  Two variants were considered as described in Table \ref{tab:estimator desc df}.

\begin{table}[!th]
	\begin{center}
		\caption{Summary of estimators compared - dual frame sample}
		\label{tab:estimator desc df}
		\begin{tabular} {p{2.5cm} p{1.7cm} p{9cm}} \toprule
			Estimator & Data Scenario & Description \\
			\toprule
				SP & Type III & The split-population estimator (\ref{eqn:Split Popn}) for $Y$ where $\hat{Y}_C$ is a Horvitz-Thompson estimator using data from $A$ \\
				SP\_Cal & Type III & The split-population estimator (\ref{eqn:Split Popn}) for $Y$, where $\hat{Y}_C$ is calibrated to population totals $(N_C,X_C)$, with $X$ being Frame Employment \\
			\bottomrule
		\end{tabular}
	\end{center}
\end{table}

Table \ref{tab:estimator desc co} describes the estimation methods applied to the cut-off probability sample.

\begin{table}[!th]
	\begin{center}
		\caption{Summary of estimators compared - cut-off sample}
		\label{tab:estimator desc co}
		\begin{tabular} {p{2.5cm} p{1.7cm} p{9cm}} \toprule
			Estimator & Data Scenario & Description \\
			\toprule
				CO+BD & Type III & A Horvitz-Thompson estimate based on the cut-off sample, added to the big data total for the small size units \\
				CO\_Cal+KWFr & Type 0 & An estimate from the cut-off sample calibrated to population totals which excludes the cut-off population, combined with an estimate based on the big data in the excluded part of the population which is weighted up by KW weights produced by linking to the frame to obtain $\delta_k, k \in B$ and frame information on $X$ \\
			\bottomrule
		\end{tabular}
	\end{center}
\end{table}

For comparison, we also produced estimates where population frame information on employment and industry were the auxiliary variables available in the estimation process.  The estimators used are described in Table \ref{tab:estimator desc bd}.

\begin{table}[!th]
	\begin{center}
		\caption{Summary of estimators compared - Big data and frame only}
		\label{tab:estimator desc bd}
		\begin{tabular} {p{2.5cm} p{1.7cm} p{9cm}} \toprule
			Estimator & Data Scenario & Description \\
			\toprule
				AuxDiv & Type I & The big data total in each Industry Division $d$ adjusted by an industry-specific factor $X_d/X_{B,d}$, and then summed over the $D$ industries to produce the overall total: $\sum_{d \in D}(X_d/X_{B,d})\sum_{k \in N_{B,d}} y_k$ \\
				KWFr & Type I & Estimation of $\hat{\pi}^B_k, k \in B$ by first linking the frame data to the big dataset to find $\delta_k, k \in B$.  A logistic regression is then fitted on the frame data to estimate $\hat{\pi}^B_k$, with frame employment and industry division as the explanatory variables in the model.  Estimates are produced using (\ref{eqn:Hajek IPW estimator}) \\
			\bottomrule
		\end{tabular}
	\end{center}
\end{table}

\subsubsection{Measurement error correction in the non-probability sample}

By adopting the measurement error model (\ref{eqn: me defn}), we can re-arrange and obtain an expression for a measurement-error corrected version of $y^*_i$:

\begin{equation} {\label{eqn:me correction}}
	\hat{y}_i = \hat{\beta}_{1}^{-1}(y_i^* - \hat{\beta}_0)
\end{equation}

In a Type III data scenario, the parameters $\beta_0$ and $\beta_1$ may be estimated using data from $A \cap B$.  In our simulation study we examined the performance of a few estimators when measurement error was present, as detailed in Table \ref{tab:estimator desc me}.

\begin{table}[!th]
	\begin{center}
		\caption{Summary of estimators - Measurement error corrected estimators}
		\label{tab:estimator desc me}
		\begin{tabular} {p{2.5cm} p{1.7cm} p{9cm}} \toprule
			Estimator & Data Scenario & Description \\
			\toprule
				KW-Cor & Type III & Apply the correction (\ref{eqn:me correction}) based on data from $A \cap B$.  Apply the model to the data in the big dataset to create $\hat{y}_k, k \in B$.  Estimates are produced by feeding these $\hat{y}_k$ into (\ref{eqn:Hajek IPW estimator}), and applying the weights from the KW estimator \\
				KW-Cal-Cor & Type III & Similar to KW-Cor, except with an additional calibration step to population size $N$ and frame employment total \\ 
				CO\_Cal+KWFr-Cor & Type III & Similar to the CO\_Cal+KWFr cut-off sample approach, but with measurement error correction for the small units.  The measurement error corrected data are then used to provide the contribution from the big dataset \\
				KWFr-Cor & Type III & Similar to the KWFr estimator in Table \ref{tab:estimator desc bd}, except now the probability sample is also utilised to fit a measurement error correction model for each data item.  Measurement error corrected versions of each data item are used instead of $y^*$ to form estimates \\
			\bottomrule
		\end{tabular}
	\end{center}
\end{table}

\subsection{Results}

Table \ref{tab:sample sizes} provides the reference sample sizes resulting from each of the different sample designs.  Sample reductions are shown relative to the single-frame design. For this simulation, the dual-frame designs provide good potential for reducing the sample size, with a saving of about 40\%.  When using the cut-off sampling, the resulting sample savings are about 11\%.  Note that we did not attempt to standardise the designs in terms of the achieved precision or RMSE of estimators, as it is unclear how such a standardisation would be defined when there are many variables and estimators being considered.  As a result, the sample sizes achieved by the different designs need to be considered in conjunction with the RMSE results achieved by those designs in Tables \ref{tab:results SAR No ME} to \ref{tab:results SNAR With ME}.

\begin{table}[!th]
	\begin{center}
		\caption{Sample sizes under different designs}
		\label{tab:sample sizes}
		\begin{tabular} {l c c } \toprule
			Sample Design & Sample Size & Sample Reduction \\
			\toprule
				Single-frame 	   & 7,715 & 0\%     \\      
				Dual-frame - SAR   & 4,559$^*$ & 41\%    \\
				Dual-frame - SNAR  & 4,598$^*$ & 40\%    \\
				Cut-off   	 	   & 6,883 & 11\%    \\
			\bottomrule
			\footnotesize{$^*$ Average sample size over all simulations}
		\end{tabular}
	\end{center}
\end{table}

In each iteration of the simulation, national level estimates were produced for four $y$ variables of interest - total weekly earnings (Earn), total reported employment (Emp), total overtime (Ovt), and average weekly earnings (AWE) defined as the ratio of total weekly earnings to total reported employment.  The RB and RRMSE were calculated for the different estimators described in Tables \ref{tab:estimator desc}, \ref{tab:estimator desc df}, \ref{tab:estimator desc co}, \ref{tab:estimator desc bd} and \ref{tab:estimator desc me}, and are shown in Tables \ref{tab:results SAR No ME} to \ref{tab:results SNAR With ME}.

Table \ref{tab:best performing estimators} lists the best-performing estimator in terms of RRMSE for each data item in each of the four scenarios of SAR vs SNAR and with vs without measurement error.  No one estimator consistently outperforms the others in all scenarios.  However, we note that the SP-Cal estimator seems to perform reasonably in the no measurement error scenario, while the GREG estimator more often performs best when there is measurement error.  We next highlight our results for each of the four classes of estimators considered.

\begin{table}[!th]
	\begin{center}
		\caption{Best-performing estimators under different scenarios}
		\label{tab:best performing estimators}
		\begin{tabular} {l l l} \toprule
			 & No Measurement Error & With Measurement Error \\
			\toprule
				\multirow{4}{*}{SAR} & Earn: SP-Cal & Earn: GREG \\
									 & Emp: KW-Cal & Emp: GREG \\
									 & Ovt: KWFr & Ovt: KWFr-Cor \\
									 & AWE: AuxDiv & AWE: KWFr \\
			\midrule
				\multirow{4}{*}{SNAR} & Earn: SP-Cal & Earn: GREG \\
									 & Emp: SP-Cal & Emp: GREG \\
									 & Ovt: ALP & Ovt: KW-Cal-Cor \\
									 & AWE: ALP & AWE: DR\_wgt \\
			\bottomrule
		\end{tabular}
	\end{center}
\end{table}

\subsubsection{Single-frame design results}

The GREG (Type 0 data structure), RDI (Type III) and QR MA (Type IV) estimators rely on the probability sample data for inferences, and as expected they have negligible relative bias under all missingness and measurement error scenarios we examined.  These estimators are ``safe'' estimators, producing robust performance.  Of these three estimators, the GREG with frame employment as the auxiliary variable performs best - it is also consistently among the best performers across all estimators.  The RDI estimator, which relies on $B$ only to provide calibration benchmarks for weighting $A$, achieves good performance, coming close to the GREG.  It is also asymptotically unbiased.  Combining the GREG and RDI benchmarks into a single calibration process can further reduce the RRMSE of the resulting estimates, however care should be observed when choosing the set of benchmarks as including too many calibration constraints may result in an increased RRMSE, or worse yet an infeasible calibration process (as noted in \cite{golini.righi_2024}).  The addition of measurement error in $y$ erodes the performance of the RDI estimator (see Tables \ref{tab:results SAR With ME} and \ref{tab:results SNAR With ME}).

\begin{table}[h!]
	\begin{center}
		\caption{Monte Carlo Bias and RRMSE of estimators based on 2000 samples - SAR, No Measurement Error}
		\label{tab:results SAR No ME}
		\begin{tabular} {c l c c c c r r r r} \toprule
			\multicolumn{2}{c}{Estimator} & \multicolumn{4}{c}{RB ($\times10^2$)} & \multicolumn{4}{c}{RRMSE ($\times10^2$)} \\
			\multicolumn{2}{c}{} & \multicolumn{1}{c}{Earn} & \multicolumn{1}{c}{Emp} & \multicolumn{1}{c}{Ovt} & \multicolumn{1}{c}{AWE} & \multicolumn{1}{c}{Earn} & \multicolumn{1}{c}{Emp} & \multicolumn{1}{c}{Ovt} & \multicolumn{1}{c}{AWE} \\
			\toprule
				\multicolumn{8}{l}{Single-frame design} \\
				& RDI   		& 0.0   & 0.0   & 0.2   & 0.0   & 1.0   & 1.0   & 6.3   & 0.4 \\
				& GREG  		& 0.0   & 0.0   & 0.2   & 0.0   & 0.8   & 0.7   & 6.3   & 0.4 \\
				& QR MA 		& 0.0   & 0.0   & 0.2   & 0.0   & 1.3   & 1.2   & 6.4   & 0.4 \\
				& KW    		& 0.1   & -0.2  & 0.3   & 0.2   & 2.0   & 2.0   & 2.4   & 0.5 \\
				& KW-Cal 		& 0.2   & 0.0   & 0.4   & 0.2   & 0.5   & 0.0   & 1.5   & 0.5 \\
				& KW-Earn 		& -0.1  & -0.3  & 0.1   & 0.2   & 2.2   & 2.2   & 2.6   & 0.5 \\
				& ALP   		& -0.2  & 0.0   & 0.6   & -0.1  & 1.1   & 1.1   & 1.3   & 0.1 \\
				& Wgt\_Reg\_MI 	& 0.2   & 0.0   & 0.6   & 0.3   & 1.2   & 1.1   & 1.3   & 0.3 \\
				& DR\_wgt 		& 0.2   & 0.0   & 0.6   & 0.3   & 1.2   & 1.1   & 1.3   & 0.3 \\
				& HD\_MI 		& 0.2   & 0.2   & 0.5   & 0.0   & 1.4   & 1.4   & 6.5   & 0.5 \\
			\midrule
				\multicolumn{8}{l}{Dual-frame design} \\
				& SP   			& 0.0   & 0.0   & 0.0   & 0.0   & 0.5   & 0.5   & 2.5   & 0.2 \\
				& SP\_Cal 		& 0.0   & 0.0   & 0.0   & 0.0   & 0.3   & 0.3   & 2.5   & 0.2 \\
			\midrule
				\multicolumn{8}{l}{Cut-off design} \\
				& CO+BD 		& -6.2  & -6.3  & -6.5  & 0.0   & 6.3   & 6.4   & 8.7   & 0.4 \\
				& CO\_Cal+KWFr  & 0.1   & 0.0   & 0.1   & 0.2   & 0.6   & 0.5   & 5.6   & 0.4 \\
			\midrule
				\multicolumn{8}{l}{Big Data only} \\
				& AuxDiv 		& -2.9  & -2.8  & -3.3  & -0.1  & 2.9   & 2.8   & 3.3   & 0.1 \\   
				& KWFr  		& 0.4   & 0.1   & 0.6   & 0.3   & 0.4   & 0.1   & 0.7   & 0.3 \\    
			\bottomrule			
		\end{tabular}
		\begin{minipage}{14cm}
			\vspace{0.1cm}
			\small Note: ``RDI'' is the calibrated estimator outlined in \citet{kim.tam_2021}; ``GREG'' is the Generalised Regression estimator; ``QR MA'' is the estimator of \citet{medous.etal_2023a}; ``KW'' and ``KW-Earn'' are IPW estimators based on \citet{kim.wang_2019}, with frame employment and log(earnings) (KW-Earn only) in the model for $\hat{\pi}^B_k$; ``KW-Cal'' is the KW estimator calibrated to $N$ and total frame employment; ``ALP'' is the \citet{wang.etal_2021} estimator; ``Wgt\_Reg\_MI'' imputes $y$ in $A$ using weighted regression models fitted on $B$; ``DR\_wgt'' is a doubly robust estimator; ``HD\_MI'' is a mass imputation for $A$ using the hot deck method; ``SP'' is the estimator (\ref{eqn:Split Popn}); ``SP\_Cal'' is (\ref{eqn:Split Popn}) with the probability sample calibrated to population totals; ``CO+BD'' is the HT estimator for the cut-off sample added to the big data total for $\mathcal{U}_E$; ``CO\_Cal+KWFr'' combines a calibrated total for $\hat{Y}_F$ and an estimate of $\hat{Y}_E$ using KW; ``AuxDiv'' is $\sum_{d \in D}(X_d/X_{B,d})\sum_{k \in N_{B,d}} y_k$; ``KWFr'' is a KW estimator with propensities estimated using frame data linked to the big dataset.
		\end{minipage}
	\end{center}
\end{table}

When there is no measurement error, the RRMSE for the QR MA estimator is slightly higher than that for the RDI estimator, reflecting a penalty due to the fact that it estimates $y_k$ for all units in the large dataset $B$, while the RDI is able to utilise the real values of $y_k, k \in B$.  When measurement error exists in $B$, however, the QR MA estimator provides slightly better RRMSE outcomes for Earn and Emp compared with the RDI which uses the mis-measured $y^*_i$ from $B$ in its benchmarks.

A few variants of propensity score estimator were examined in our study, applicable under a Type II data structure.  The KW-Cal estimator performs very well in the ideal scenario of SAR and no measurement error (see Table \ref{tab:results SAR No ME}).  In general, the calibration to frame employment helps to reduce the variance of the KW estimates.

The ALP estimator performs favourably as an alternative to the KW estimator, achieving a lower RRMSE in the four scenarios tested.  In our simulation study, we found that some of the resulting propensities were larger than 1 when using this method, and this lead to IPW weights of less than 1.  This, along with other conceptual issues with the pooled approach noted by \citet{wu_2022a}, means the survey practitioner will need to consider how applicable this approach will be for their case.  In our study, though, these issues did not hinder the effectiveness of the estimator to produce relatively efficient, unbiased estimates of population totals.

In the SAR case, the inclusion of earnings in the KW-Earn estimator did not produce reductions in RRMSE over the KW estimator, which is expected since in the SAR case missingness in $B$ does not depend on earnings (see Table \ref{tab:results SAR No ME}).  On the other hand, in the SNAR case where the propensity of selection into $B$ is also influenced by earnings, the inclusion of earnings in the model yielded lower RB and RRMSE (see Table \ref{tab:results SNAR No ME}) compared with the KW estimator.

The performance of the mass imputation approach Wgt\_Reg\_MI in our study was better than the KW estimator in all scenarios, although both suffered under measurement error.  The inclusion of estimated propensity weights in the regression imputation model improved the performance of the regression model, aligning with findings in \citet{castro-martin.etal_2022}.  The DR\_wgt estimator, which provides protection against mis-specification in one of the IPW or Regression models, tends to have comparable RRMSE compared with the Wgt\_Reg\_MI estimator.

The HD\_MI estimator is a non-parametric mass imputation approach for $A$.  This estimator was chosen as a less computationally intensive approach (and hence faster) compared with k Nearest Neighbour imputation.  Compared to the parametric regression mass imputation approach, HD\_MI produced comparable results for 3 of the four data items of interest in all scenarios except the ideal SAR without measurement error case.  The Ovt data item was the one item where the hot deck imputation did not perform as well as the regression model.  The results suggest that, when suitable imputation classes are formed using the covariates $\boldsymbol{x}$, the hot-deck method may be a faster, simpler to implement alternative to nearest neighbour imputation.

\begin{table}[h!]
	\begin{center}
		\caption{Monte Carlo Bias and RRMSE of estimators based on 2000 samples - SAR, With Measurement Error }
		\label{tab:results SAR With ME}
		\begin{tabular} {c l c c c c r r r r} \toprule
			\multicolumn{2}{c}{Estimator} & \multicolumn{4}{c}{RB ($\times10^2$)} & \multicolumn{4}{c}{RRMSE ($\times10^2$)} \\
			\multicolumn{2}{c}{} & \multicolumn{1}{c}{Earn} & \multicolumn{1}{c}{Emp} & \multicolumn{1}{c}{Ovt} & \multicolumn{1}{c}{AWE} & \multicolumn{1}{c}{Earn} & \multicolumn{1}{c}{Emp} & \multicolumn{1}{c}{Ovt} & \multicolumn{1}{c}{AWE} \\
			\toprule
				\multicolumn{8}{l}{Single-frame design} \\
				& RDI   		& 0.2   & 0.1   & 0.3   & 0.0   & 1.4   & 1.3   & 6.4   & 0.4 \\
				& GREG  		& 0.0   & 0.0   & 0.2   & 0.0   & 0.8   & 0.7   & 6.3   & 0.4 \\
				& QR MA 		& 0.0   & 0.0   & 0.2   & 0.0   & 1.3   & 1.2   & 6.4   & 0.4 \\
				& KW    		& -12.9 & -12.8 & -13.0 & -0.2  & 13.0  & 12.9  & 13.2  & 0.5 \\
				& KW-Cal 		& -12.8 & -12.6 & -12.9 & -0.2  & 12.8  & 12.6  & 13.0  & 0.5 \\
				& KW-Cor 		& -0.1  & -0.4  & -0.1  & 0.3   & 3.2   & 3.6   & 5.3   & 1.2 \\
				& KW-Cal-Cor 	& 0.1   & -0.3  & 0.1   & 0.3   & 2.5   & 3.0   & 4.8   & 1.2 \\
				& KW-Earn 		& -13.0 & -12.9 & -13.1 & -0.2  & 13.2  & 13.1  & 13.3  & 0.5 \\
				& ALP   		& -12.9 & -12.4 & -12.5 & -0.6  & 13.0  & 12.5  & 12.6  & 0.6 \\
				& Wgt\_Reg\_MI 	& -12.8 & -12.6 & -12.8 & -0.1  & 12.8  & 12.7  & 12.8  & 0.2 \\
				& DR\_wgt 		& -12.8 & -12.6 & -12.8 & -0.1  & 12.8  & 12.7  & 12.8  & 0.2 \\
				& HD\_MI 		& -12.8 & -12.4 & -12.7 & -0.4  & 12.9  & 12.6  & 14.8  & 1.0 \\
			\midrule
				\multicolumn{8}{l}{Dual-frame design} \\
				& SP    		& -8.2  & -7.9  & -8.4  & -0.4  & 8.3   & 7.9   & 8.8   & 0.4 \\
				& SP\_Cal 		& -8.2  & -7.9  & -8.4  & -0.4  & 8.2   & 7.9   & 8.8   & 0.4 \\
			\midrule
				\multicolumn{8}{l}{Cut-off design} \\
				& CO+BD 			& -7.2  & -7.1  & -7.5  & -0.1  & 7.3   & 7.2   & 9.4   & 0.4 \\
				& CO\_Cal+KWFr 		& -1.7  & -1.7  & -1.8  & 0.0   & 1.9   & 1.8   & 5.9   & 0.4 \\
				& CO\_Cal+KWFr-Cor 	& 0.2   & 0.1   & 0.3   & 0.0   & 0.9   & 0.8   & 5.7   & 0.6 \\
			\midrule
				\multicolumn{8}{l}{Big Data only} \\
				& AuxDiv 		& -15.1 & -14.7 & -15.8 & -0.4  & 15.1  & 14.7  & 15.8  & 0.5 \\
				& KWFr  		& -12.7 & -12.5 & -12.7 & -0.1  & 12.7  & 12.5  & 12.7  & 0.2 \\
				& KWFr-Cor 		& 0.2   & -0.1  & 0.3   & 0.3   & 2.5   & 2.9   & 4.6   & 1.1 \\
			\bottomrule
		\end{tabular}
		\begin{minipage}{14cm}
			\vspace{0.1cm}
			\small Note: ``RDI'' is the calibrated estimator outlined in \citet{kim.tam_2021}; ``GREG'' is the Generalised Regression estimator; ``QR MA'' is the estimator of \citet{medous.etal_2023a}; ``KW'' and ``KW-Earn'' are IPW estimators based on \citet{kim.wang_2019}, with frame employment and log(earnings) (KW-Earn only) in the model for $\hat{\pi}^B_k$; ``KW-Cal'' is the KW estimator calibrated to $N$ and total frame employment; ``KW-Cor'' and ``KW-Cal-Cor'' are measurement error corrected versions of ``KW'' and ``KW-Cal''; ``ALP'' is the \citet{wang.etal_2021} estimator; ``Wgt\_Reg\_MI'' imputes $y$ in $A$ using weighted regression models fitted on $B$; ``DR\_wgt'' is a doubly robust estimator; ``HD\_MI'' is a mass imputation for $A$ using the hot deck method; ``SP'' is the estimator (\ref{eqn:Split Popn}); ``SP\_Cal'' is (\ref{eqn:Split Popn}) with the probability sample calibrated to population totals; ``CO+BD'' is the HT estimator for the cut-off sample added to the big data total for $\mathcal{U}_E$; ``CO\_Cal+KWFr'' combines a calibrated total for $\hat{Y}_F$ and an estimate of $\hat{Y}_E$ using KW; ``CO\_Cal+KWFr-Cor'' is a measurement error corrected version of ``CO\_Cal+KWFr-Cor''; ``AuxDiv'' is $\sum_{d \in D}(X_d/X_{B,d})\sum_{k \in N_{B,d}} y_k$; ``KWFr'' is a KW estimator with propensities estimated using frame data linked to the big dataset; ``KWFr-Cor'' is a measurement error corrected version of ``KWFr''.
		\end{minipage}
	\end{center}
\end{table}

It is worth pointing out the RRMSE results for the Overtime variable when no measurement error is present (Tables \ref{tab:results SAR No ME} and \ref{tab:results SNAR No ME}).  The Overtime variable has a larger population variance and lower correlation with the benchmarking variable Frame Employment compared with Earnings and Reported Employment.  We did not include the total value of Overtime in $B$ as a calibration constraint for the RDI and QR MA estimators, and as a result the RRMSE results for the RDI and QR MA estimators tend to be large.  We could of course include the $B$ totals for all the variables of interest in the calibration, but as noted above this may lead to more variable weights and hence higher RRMSE, or an infeasible carlibation process.  This suggests that in a multi-purpose survey with a potentially large number of disparate variables, these estimators may not be suitable as it will be infeasible to include all of the different variables in the calibration process.

More generally for Overtime when no measurement error is present, the estimators based on the probability sample $A$ - RDI, GREG, QR MA and HD\_MI - tend to have a large RRMSE.  On the other hand, the KW-based, ALP, Wgt\_Reg\_MI and DR\_wgt methods achieve a significantly lower RRMSE; these estimators all base their inference on the much larger dataset $B$ which helps in lowering the variance of the Overtime estimates.

From Tables \ref{tab:results SAR With ME} and \ref{tab:results SNAR With ME}, the KW-based, HD\_MI, Wgt\_Reg\_MI and DR\_wgt approaches applicable under a Type II data structure all yield poor results under measurement error due to their reliance on the data in $B$.  This was true under both a SAR and SNAR setting.  None of these approaches included any provision for erroneously measured data items.  Measurement error correction successfully negates the bias from measurement error when the measurement error model holds.  However, the cost is a much higher overall RRMSE.

\subsubsection{Dual-frame design results}

In the without measurement error scenarios, the split-population estimators are unbiased, and their RRMSE results are generally among the lowest of all estimators we examined across SAR and SNAR settings (see Tables \ref{tab:results SAR No ME} and \ref{tab:results SNAR No ME}).  Including a calibration to auxiliary population totals for the dual-frame estimator helps to reduce RRMSE even further for earnings and reported employment.

However, measurement error in $B$ leads to bias in the split-population estimators due to the reliance on $B$ to produce a value for $Y_B$ (see Tables \ref{tab:results SAR With ME} and \ref{tab:results SNAR With ME}).  The dual-frame samples here don't have any overlap between $A$ and $B$.  Including some overlap between the samples $A$ and $B$ may be desirable to provide some data for correcting the measurement error.

\subsubsection{Cut-off design results}

The results for the estimators using the cut-off sample design show that in our case simply combining the reference sample and the portion of $B$ under the cut-off threshold did not account for all the contribution below the threshold.  This was because there was a significant negative bias due to the contribution of units under the cut-off threshold which were also not in $B$.  This was exacerbated by the fact that small units are less likely to be in $B$.

\begin{table}[h!]
	\begin{center}
		\caption{Monte Carlo Bias and RRMSE of estimators based on 2000 samples - SNAR, No Measurement Error }
		\label{tab:results SNAR No ME}
		\begin{tabular} {c l c c c c r r r r} \toprule
			\multicolumn{2}{c}{Estimator} & \multicolumn{4}{c}{RB ($\times10^2$)} & \multicolumn{4}{c}{RRMSE ($\times10^2$)} \\
			\multicolumn{2}{c}{} & \multicolumn{1}{c}{Earn} & \multicolumn{1}{c}{Emp} & \multicolumn{1}{c}{Ovt} & \multicolumn{1}{c}{AWE} & \multicolumn{1}{c}{Earn} & \multicolumn{1}{c}{Emp} & \multicolumn{1}{c}{Ovt} & \multicolumn{1}{c}{AWE} \\
			\toprule
				\multicolumn{8}{l}{Single-frame design} \\
				& RDI   		& 0.0   & 0.0   & -0.1  & 0.0   & 1.1   & 1.0   & 6.5   & 0.4 \\
				& GREG  		& 0.0   & 0.0   & -0.1  & 0.0   & 0.8   & 0.7   & 6.4   & 0.4 \\
				& QR MA 		& 0.0   & 0.0   & -0.1  & 0.0   & 1.3   & 1.2   & 6.5   & 0.4 \\
				& KW    		& -3.0  & -3.1  & -2.9  & 0.1   & 3.6   & 3.6   & 3.9   & 0.5 \\
				& KW-Cal 		& -1.5  & -1.6  & -1.6  & 0.1   & 1.6   & 1.6   & 2.3   & 0.5 \\
				& KW-Earn 		& -0.2  & -0.2  & -0.1  & 0.0   & 1.9   & 1.8   & 2.5   & 0.5 \\
				& ALP   		& -1.7  & -1.6  & -1.5  & -0.1  & 2.1   & 2.0   & 1.9   & 0.1 \\
				& Wgt\_Reg\_MI 	& -1.4  & -1.6  & -1.5  & 0.2   & 1.9   & 2.0   & 1.9   & 0.2 \\
				& DR\_wgt 		& -1.4  & -1.6  & -1.5  & 0.2   & 1.8   & 1.9   & 1.9   & 0.2 \\
				& HD\_MI 		& -1.3  & -1.2  & -1.2  & -0.1  & 1.9   & 1.8   & 6.8   & 0.5 \\
			\midrule
				\multicolumn{8}{l}{Dual-frame design} \\
				& SP    		& 0.0   & 0.0   & 0.0   & 0.0   & 0.6   & 0.6   & 2.9   & 0.2 \\
				& SP\_Cal 		& 0.0   & 0.0   & 0.0   & 0.0   & 0.4   & 0.3   & 2.8   & 0.2 \\
			\midrule
				\multicolumn{8}{l}{Cut-off design} \\
				& CO+BD 		& -6.6  & -6.6  & -6.8  & 0.0   & 6.7   & 6.7   & 8.9   & 0.4 \\
				& CO\_Cal+KWFr 	& -0.8  & -0.9  & -0.8  & 0.1   & 1.0   & 1.0   & 5.7   & 0.4 \\
			\midrule
				\multicolumn{8}{l}{Big Data only} \\
				& AuxDiv 		& -3.3  & -3.2  & -3.9  & -0.2  & 3.3   & 3.2   & 4.0   & 0.2 \\  
				& KWFr  		& -2.7  & -2.8  & -2.6  & 0.1   & 2.7   & 2.8   & 2.7   & 0.1 \\  
			\bottomrule
		\end{tabular}
		\begin{minipage}{14cm}
			\vspace{0.1cm}
			\small Note: ``RDI'' is the calibrated estimator outlined in \citet{kim.tam_2021}; ``GREG'' is the Generalised Regression estimator; ``QR MA'' is the estimator of \citet{medous.etal_2023a}; ``KW'' and ``KW-Earn'' are IPW estimators based on \citet{kim.wang_2019}, with frame employment and log(earnings) (KW-Earn only) in the model for $\hat{\pi}^B_k$; ``KW-Cal'' is the KW estimator calibrated to $N$ and total frame employment; ``ALP'' is the \citet{wang.etal_2021} estimator; ``Wgt\_Reg\_MI'' imputes $y$ in $A$ using weighted regression models fitted on $B$; ``DR\_wgt'' is a doubly robust estimator; ``HD\_MI'' is a mass imputation for $A$ using the hot deck method; ``SP'' is the estimator (\ref{eqn:Split Popn});  ``SP\_Cal'' is (\ref{eqn:Split Popn}) with the probability sample calibrated to population totals; ``CO+BD'' is the HT estimator for the cut-off sample added to the big data total for $\mathcal{U}_E$; ``CO\_Cal+KWFr'' combines a calibrated total for $\hat{Y}_F$ and an estimate of $\hat{Y}_E$ using KW; ``AuxDiv'' is $\sum_{d \in D}(X_d/X_{B,d})\sum_{k \in N_{B,d}} y_k$; ``KWFr'' is a KW estimator with propensities estimated using frame data linked to the big dataset.
		\end{minipage}
	\end{center}
\end{table}

The CO\_Cal+KWFr estimator aims to account explicitly for the contribution of the excluded part of the population $\mathcal{U}_E$ using the KW estimator and available frame information.  In the without-measurement error scenarios, this effectively corrects for the selection bias, and also results in a low RRMSE for all items except overtime.  When measurement error exists, using the big data to account for the contribution of $\mathcal{U}_E$ means that there will be some bias in the estimates due to measurement error.  The estimator CO\_Cal+KWFr-Cor which applies a measurement error model using available auxiliary information (in this case, from the frame) assists in reducing the RB, with an overall RRMSE that ranks well compared with the other estimators.

In the SAR without measurement error scenario (Table \ref{tab:results SAR No ME}) there is little reason to opt for the CO\_Cal+KWFr estimator rather than the KWFr estimator.  In this case, where the quality of the data in $B$ is reliable, the reference sample is less necessary for producing efficient estimates.  However, when SNAR or measurement error exists, the reference sample provides a safety net, providing a means to help correct the measurement error and reduce the impact from SNAR missingness in the big data.

For the worst-case SNAR with measurement error scenario, the CO\_Cal+KWFr-Cor estimator yielded close to the lowest overall RRMSE for the data items of interest (see Table \ref{tab:results SNAR With ME}).  This was the case even though the contribution from the KW estimator did not fully reflect the non-response mechanism since that estimator did not include earnings (hence there is still a negative bias in the estimates obtained).  This demonstrates the benefit of not relying solely on $B$ to produce estimates (such as in the single-frame KW-based estimators).  Additionally, the exclusion of the smallest units from the cut-off sample helps to reduce the overall sample variance.

\begin{table}[h!]
	\begin{center}
		\caption{Monte Carlo Bias and RRMSE of estimators based on 2000 samples - SNAR, With Measurement Error }
		\label{tab:results SNAR With ME}
		\begin{tabular} {c l c c c c r r r r} \toprule
			\multicolumn{2}{c}{Estimator} & \multicolumn{4}{c}{RB ($\times10^2$)} & \multicolumn{4}{c}{RRMSE ($\times10^2$)} \\
			\multicolumn{2}{c}{} & \multicolumn{1}{c}{Earn} & \multicolumn{1}{c}{Emp} & \multicolumn{1}{c}{Ovt} & \multicolumn{1}{c}{AWE} & \multicolumn{1}{c}{Earn} & \multicolumn{1}{c}{Emp} & \multicolumn{1}{c}{Ovt} & \multicolumn{1}{c}{AWE} \\
			\toprule
				\multicolumn{8}{l}{Single-frame design} \\
				& RDI   		& 0.2   & 0.2   & 0.1   & 0.0   & 1.4   & 1.3   & 6.5   & 0.4 \\
				& GREG  		& 0.0   & 0.0   & -0.1  & 0.0   & 0.8   & 0.7   & 6.4   & 0.4 \\
				& QR MA 		& 0.0   & 0.0   & -0.1  & 0.0   & 1.3   & 1.2   & 6.5   & 0.4 \\
				& KW    		& -15.4 & -15.1 & -15.6 & -0.3  & 15.5  & 15.2  & 15.8  & 0.6 \\
				& KW-Cal 		& -14.1 & -13.8 & -14.5 & -0.3  & 14.1  & 13.8  & 14.6  & 0.5 \\
				& KW-Cor 		& -3.0  & -3.2  & -3.1  & 0.2   & 4.4   & 4.8   & 6.3   & 1.4 \\
				& KW-Cal-Cor 	& -1.5  & -1.7  & -1.8  & 0.2   & 2.9   & 3.5   & 5.4   & 1.4 \\
				& KW-Earn 		& -13.1 & -12.8 & -13.4 & -0.4  & 13.2  & 12.9  & 13.6  & 0.6 \\
				& ALP   		& -14.2 & -13.7 & -14.2 & -0.5  & 14.2  & 13.8  & 14.3  & 0.5 \\
				& Wgt\_Reg\_MI 	& -14.0 & -13.8 & -14.4 & -0.3  & 14.1  & 13.8  & 14.4  & 0.3 \\
				& DR\_wgt		& -14.0 & -13.8 & -14.4 & -0.3  & 14.1  & 13.8  & 14.4  & 0.3 \\
				& HD\_MI 		& -14.0 & -13.6 & -14.2 & -0.5  & 14.2  & 13.8  & 16.0  & 1.0 \\
			\midrule
				\multicolumn{8}{l}{Dual-frame design} \\
				& SP    		& -7.6  & -7.3  & -7.7  & -0.3  & 7.6   & 7.3   & 8.2   & 0.4 \\
				& SP\_Cal 		& -7.6  & -7.3  & -7.7  & -0.3  & 7.6   & 7.3   & 8.2   & 0.4 \\
			\midrule
				\multicolumn{8}{l}{Cut-off design} \\
				& CO+BD 			& -7.5  & -7.4  & -7.7  & -0.1  & 7.6   & 7.5   & 9.6   & 0.4 \\
				& CO\_Cal+KWFr 		& -2.5  & -2.5  & -2.6  & 0.0   & 2.6   & 2.6   & 6.2   & 0.4 \\
				& CO\_Cal+KWFr-Cor 	& -0.7  & -0.8  & -0.6  & 0.1   & 1.2   & 1.2   & 5.8   & 0.6 \\
			\midrule
				\multicolumn{8}{l}{Big Data only} \\
				& AuxDiv 		& -15.5 & -15.0 & -16.4 & -0.6  & 15.5  & 15.0  & 16.4  & 0.6 \\
				& KWFr  		& -15.1 & -14.9 & -15.4 & -0.3  & 15.1  & 14.9  & 15.4  & 0.3 \\
				& KWFr-Cor 		& -2.7  & -2.9  & -2.8  & 0.2   & 3.6   & 4.2   & 5.6   & 1.3 \\ 
			\bottomrule
		\end{tabular}
		\begin{minipage}{14cm}
			\vspace{0.1cm}
			\small Note: ``RDI'' is the calibrated estimator outlined in \citet{kim.tam_2021}; ``GREG'' is the Generalised Regression estimator; ``QR MA'' is the estimator of \citet{medous.etal_2023a}; ``KW'' and ``KW-Earn'' are IPW estimators based on \citet{kim.wang_2019}, with frame employment and log(earnings) (KW-Earn only) in the model for $\hat{\pi}^B_k$; ``KW-Cal'' is the KW estimator calibrated to $N$ and total frame employment; ``KW-Cor'' and ``KW-Cal-Cor'' are measurement error corrected versions of ``KW'' and ``KW-Cal''; ``ALP'' is the \citet{wang.etal_2021} estimator; ``Wgt\_Reg\_MI'' imputes $y$ in $A$ using weighted regression models fitted on $B$; ``DR\_wgt'' is a doubly robust estimator; ``HD\_MI'' is a mass imputation for $A$ using the hot deck method; ``SP'' is the estimator (\ref{eqn:Split Popn}); ``SP\_Cal'' is (\ref{eqn:Split Popn}) with the probability sample calibrated to population totals; ``CO+BD'' is the HT estimator for the cut-off sample added to the big data total for $\mathcal{U}_E$; ``CO\_Cal+KWFr'' combines a calibrated total for $\hat{Y}_F$ and an estimate of $\hat{Y}_E$ using KW; ``CO\_Cal+KWFr-Cor'' is a measurement error corrected version of ``CO\_Cal+KWFr-Cor''; ``AuxDiv'' is $\sum_{d \in D}(X_d/X_{B,d})\sum_{k \in N_{B,d}} y_k$; ``KWFr'' is a KW estimator with propensities estimated using frame data linked to the big dataset; ``KWFr-Cor'' is a measurement error corrected version of ``KWFr''.
		\end{minipage}
	\end{center}
\end{table}

\subsubsection{Big Data only results}

When aggregate auxiliary information is available from the population it can be used to help reduce selection bias, as evidenced by the performance of the AuxDiv estimator.  The AuxDiv estimator does not remove all bias as it does not include all of the $\boldsymbol{x}$ variables related to the probability of inclusion in $B$.  The presence of measurement error increases the bias of the estimate.

The KWFr estimator performs very well in the SAR without measurement error scenario (see Table \ref{tab:results SAR No ME}), achieving close to the lowest RRMSE for all data items.  This estimator uses unit-level auxiliary information for the full population to estimate the propensity scores for the large dataset $B$.  Most of the contribution to RRMSE comes from bias rather than variance.

Measurement error degrades the performance of the KWFr estimator (see Tables \ref{tab:results SAR With ME} and \ref{tab:results SNAR With ME}).  Similarly, in the SNAR scenarios the estimator does not perform as well since it is not able to include the earnings variable in the propensity model.  The results for the KWFr and AuxDiv estimators demonstrate that relying on the non-probability dataset itself for inference are likely to lead to biased estimates if there are inherent issues with either the measurement of its data items or incompleteness of available auxiliary information from the poopulation frame.

\section{Concluding remarks} \label{conclusion}

A variety of approaches, ranging from weighting-based methods, to model-based or imputation approaches, to combinations of the two, have been developed to address the faults of non-probability data.  The objective of this paper is to compare how a range of these methods perform in business survey context assuming different missingness and measurement error settings for a large non-probability dataset, and a reference probability sample available to assist.  The results in the paper provide valuable insight into the usefulness of these methods under various conditions, and are important for increasing the efficiency of statistics produced while reducing respondent burden and sample sizes.

When auxiliary information, related to $y$ or which effectively describe the missingness mechanism in $B$, is available and used in the estimation process, the methods we examined can effectively account for selection bias.  In the most ideal scenario of SAR missingness and no measurement error in the non-probability dataset, it is not imperative to have a reference sample which overlaps with the non-probability data.  When the non-probability data is big and can be linked to the population frame the use of a calibrated split population estimator provides a beneficial combination of a low reference sample size and the best accuracy of the estimators we considered.  The KW estimator calibrated to population totals also performs very well in this scenario, assuming the working model for $\pi^B_i$ holds.

When there is SNAR missingness in the non-probability dataset but no measurement error, the calibrated split population estimator still provides the best results.  This approach is robust to the selection mechanism at play in the non-probability dataset, since we use the non-probability data as-is and supplement it with data from the reference sample to cover the contribution for the population not in the non-probability dataset.  The estimator is more efficient for large non-probability datasets.  In situations when the non-probability dataset is a small fraction of the population, for example a small web panel survey, the performance of the estimator will be closer to the GREG.

The presence of measurement error in the non-probability data source affects the performance of the estimators, such that the best estimator tended not to heavily rely on the non-probability data.  In our study, the GREG, RDI and QR MA estimators - all of which rely on the reference sample as the basis for estimation - tended to perform the best in the with measurement error scenarios.  An advantage of the RDI estimator over the measurement error corrected estimators listed in Table \ref{tab:estimator desc me} is that we can continue to use the $y$ values in $B$ as-is - the form of the estimator does not need to change, and remains unbiased.

The corrected cut-off estimator, which combines the cut-off sample contribution with an estimate for the excluded part of the population, also performs well under measurement error.  Further research could be beneficial for determining appropriate cut-off thresholds for the reference sample which take advantage of the ability to model or estimate the contribution from the excluded part of the population based on the available non-probability data.

Collecting information on the data item(s) of interest in the probability sample would assist with developing a model to correct for measurement error.  One factor to be wary of is that the measurement error model may introduce additional variability into the estimates, so that we may choose instead to use the probability sample as the basis for inference and not rely on the non-probability data.

Although the RDI estimator is never the best-performing estimator, it is (asymptotically) unbiased, and has a reasonably low RRMSE in all scenarios.  The estimator is also robust to SNAR situations as well as the presence of measurement error, but its efficiency can suffer under those less-than-ideal scenarios.  When data for $y$ is available in both $A$ and $B$ (Type III data structure) then the use of the RDI estimator, with the addition of population level benchmarks in the calibration, could be a relatively safe, low-risk approach which yields some gains while potentially not maximising them.  We note the need to ensure the total number of benchmarks applied does not become excessively large.  One could also combine the RDI estimator with a split-population approach in the design of $A$ to obtain a more efficient reference sample.

Many of the methods that have been developed make two assumptions: \textit{ignorability}, which implies a SAR scenario in the non-probability data source, and \textit{common support}, which implies that the probability of being in the non-probability dataset $B$ is non-zero for $i \in \mathcal{U}$.  These assumptions may not necessarily hold.  In the SNAR scenario, it may not be appropriate to just include all available data items in the propensity model.  Methods are being developed to address the SNAR missingness scenario - see for example \citet{marella_2023} and \citet{kim.morikawa_2023}.  \citet{chen.etal_2023} consider approaches to estimation when the common support assumption is violated and some part of the population does not have any chance of being selected into the non-probability sample.  This may occur, for instance, with some social media datasets.  Further development of methods that require fewer assumptions to be effective is an area for future research.

Including all $y$ data items in the propensity score model is not necessarily helpful as a catch-all approach to include potentially helpful covariates to explain SNAR non-response.  This begs the question: How can we test whether to use an approach based on SAR or one that assumes a SNAR situation, and how do we determine which $y$ data item(s) need to be included in the SNAR model? These questions will be explored further by the authors.  \citeauthor{meng_2018} (\citeyear{meng_2018}, \citeyear{meng_2022}) has put forward the notion of \textit{data defect correlation} and suggested to miniaturise this quantity to eliminate bias.  Work such as that conducted by \citet{andridge.etal_2019} on evaluating potential selection bias due to non-ignorable selection may be another avenue to pursue.

\section*{Acknowledgements}

The authors are grateful to three anonymous referees and an associate editor for their constructive comments, which have improved this article greatly.  The authors would like to thank Dr. Siu-Ming Tam and Dr. Ryan Covey for their constructive comments on an early draft of this manuscript.  The first author benefited from email exchanges with the authors of some of the methods investigated in this paper, including Prof. Yan Li and Prof. Pengfei Li.  The first author was supported by funding from the Sir Roland Wilson Foundation and the Australian Bureau of Statistics.

\section*{Disclaimer}

The views expressed in this paper are those of the authors and do not necessarily represent the views of the Australian Bureau of Statistics.  Where quoted or used, they should be attributed clearly to the authors.

\newpage
\section*{Supplemental Data}

\appendixpageoff
\appendixtitleoff
\renewcommand{\appendixtocname}{Supplemental Data}

\begin{appendices}

\setcounter{table}{0}
\crefalias{section}{supp}

\renewcommand{\thetable}{B\arabic{table}}

\section{Generating the population} \label{appendix:generate population}

\textit{Step 1: Derive input probabilities}

Data from the ABS publication \textit{8165.0 Counts of Australian Businesses, including Entries and Exits} provides counts of the number of businesses in each State by 4-digit Australia and New Zealand Standard Industry Classification (ANZSIC) code by broad employment size group.  We aggregated the business counts to the ANZSIC Division level.  Employee tax data (\cite{businesslongitudinalanalysisdataenvironmentblade_2020}) for the 2019/2020 financial year was obtained from the ABS DataLab, and the business counts from this dataset were used to apportion the broad employment counts into finer size groups.  Business counts were then translated into proportions: i) the proportion of businesses in each size group, ii) the proportion of businesses within a size group belonging to a particular industry, and iii) the proportion of businesses within a size group and industry that belong to a particular state.

\textit{Step 2: Generate frame and reported employment}

The frame and reported employment variables are assumed to have some relationship with each other, while also individually exhibiting certain characteristics in terms of their mean, variance, skewness and kurtosis.  The method described in \citet{vale.maurelli_1983} (from here on referred to as Vale-Maurelli) was used to generate data for these two variables.  Data was generated separately for each industry within a size group.  The number of records generated in each industry and size group was based on the overall population size $N$ and the proportions generated in Step 1.  The parameters fed into the Vale-Maurelli method for each size group are provided in Table \ref{tab:vm parms}.

State indicators were assigned to the generated data randomly based on a multinomial distribution, with probabilities based on the state probabilities calculated in Step 1.

\textit{Step 3: Generate total wages and salaries}

Data from the November 2021 cycle of the ABS publication \textit{Average Weekly Earnings, Australia} (\cite{australianbureauofstatistics_2022b}) was used to inform our modelling of wages and salaries data.  Within each industry division and size group, a total wages figure for the $i$'th record was generated as

\[ y_i=1740f_{id}x_{i,rep}+\epsilon \]

where $f_{id}=w_d+\text{Normal}(0,v_s)$ is a randomised wage inflation factor for the $d$'th industry division relative to the Australian-level average earnings (\$1,740) based on industry wage factors $w_d$ sourced from the AWE publication, $v_s$ is an error variance term for size group $s$, $x_{i,rep}$ is the simulated reported employment for the $i$'th business, and $\epsilon=\text{Normal}(0,v_ex_i)$ is a random ``error'' for the wages and salaries figure, normally distributed with variance increasing as $x_i$ increases.  The term $v_e$ is a multiplier which varies depending on size group.

The values used to generate total wages and salaries can be found in Tables \ref{tab:industry wage factors} and \ref{tab:error variance}.

\textit{Step 4: Generate overtime payments}

A mixture model was developed to produce the overtime payment data item.  A bernoulli distribution was firstly used to model the likelihood of non-zero overtime payments in each industry.  Different probabilities were used for each industry, ranging from 0.1 to 0.5.  For businesses with a non-zero overtime payment, an exponential model with mean 0.1 was used to derive a factor $f$ denoting the magnitude of overtime payments as a proportion of total earnings.  The total overtime payment was derived as

\[ y_{\text{overtime}} = fy_{\text{earnings}} \]

\textit{Step 5: Generate data items containing measurement error}

A measurement error version of total earnings, $y_{\text{earnings}}^*$, was produced using a two-step process.  First, the original/true earnings value for each unit was perturbed by a Normal(0.85,0.025) factor.  After this, an additional, larger contamination was added to a random subsample of the population.

Measurement error versions of reported employment and overtime were then created by multiplying the original/true values by the factor $y_{\text{earnings}}^* / y_\text{earnings}$.

\setcounter{table}{0}
\renewcommand{\thetable}{B\arabic{table}}

\section{Parameters used generate simulated data} \label{appendix: parms}

Table \ref{tab:vm parms} provides the parameter values used in the Vale-Maurelli process to generate the frame employment and reported employment data items.  Separate data generation processes were run for each of 14 size groups.

\begin{table}[H]
	\begin{center}
		\caption{Parameter values fed into Vale-Maurelli process - Frame Employment}
		\label{tab:vm parms}
		\begin{tabular} {l r r r r r r r r r} \toprule
			\multicolumn{1}{c}{Emp Group} & \multicolumn{4}{c}{Frame Employment} & \multicolumn{4}{c}{Reported Employment} & \multicolumn{1}{c}{Covariance} \\
			\multicolumn{1}{c}{} & \multicolumn{1}{c}{Mean} & \multicolumn{1}{c}{Var} & \multicolumn{1}{c}{Skew} & \multicolumn{1}{c}{Kurt} & \multicolumn{1}{c}{Mean} & \multicolumn{1}{c}{Var} & \multicolumn{1}{c}{Skew} & \multicolumn{1}{c}{Kurt} & \multicolumn{1}{c}{} \\
			\toprule
				0-4     & 2 & 3 & 1.1 & 1.2 & 3.5 & 10 & 1.2 & 1.4 & 4 \\      
				5-19    & 8 & 4.5 & 1.1 & 2 & 8.5 & 10 & 1.2 & 1.8 & 4 \\
				20-49   & 26 & 28 & 0.8 & 1.8 & 25 & 35 & 0.9 & 1.6 & 25\\
				50-99   & 68 & 60 & 0.6 & 3.55 & 51 & 70 & 0.9 & 3.6 & 55 \\
				100-149 & 120 & 70 & 0.6 & 2.2 & 110 & 70 & 0.8 & 2.4 & 55 \\				
				150-199 & 160 & 100 & 0.4 & 1.6 & 145 & 120 & 0.7 & 1.9 & 70 \\
				200-249 & 220 & 2,200 & 0.5 & 1.5 & 200 & 2,400 & 0.5 & 1.3 & 2,000 \\
				249-299 & 280 & 6,300 & 0.7 & 1.2 & 250 & 5,400 & 0.7 & 1.5 & 4,800 \\
				300-349 & 320 & 14,000 & 0.7 & 1.2 & 270 & 16,000 & 0.9 & 1.3 & 12,000 \\
				349-399 & 375 & 25,000 & 0.8 & 0.8 & 300 & 25,000 & 0.9 & 1.1 & 21,000 \\
				400-449 & 420 & 45,000 & 1.03 & 1.2 & 370 & 45,000 & 1.02 & 1.1 & 35,000 \\
				449-499 & 470 & 58,000 & 1.1 & 1.3 & 420 & 55,000 & 1.1 & 1.5 & 47,000 \\
				500-999 & 650 & 80,000 & 0.85 & 0.8 & 550 & 72,000 & 0.8 & 0.8 & 66,000 \\
				1000+  & 1,350 & 200,000 & 2.3 & 9 & 1,200 & 200,000 & 2 & 8 & 190,000 \\
			\bottomrule
		\end{tabular}
		\begin{minipage}{16cm}
			\vspace{0.1cm}
			\small Note: Parameters created based on Pay-As-You-Go (PAYG) data from the 2019-2020 financial year (\cite{businesslongitudinalanalysisdataenvironmentblade_2020}), merged on to Economic Activity Survey data from the 2020-2021 financial year (\cite{australianbureauofstatistics_2021}), accessed through the ABS DataLab 2 March, 2023.
		\end{minipage}
	\end{center}
\end{table} 

\section{Factors used to generate wages and salaries}

\setcounter{table}{0}
\renewcommand{\thetable}{C\arabic{table}}

Table \ref{tab:industry wage factors} provides the wage factors used for each industry.  The factors represent the relative magnitude of each industry's weekly wage compared with the national average.

\begin{table}[H]
	\begin{center}
		\caption{Industry Wage Factors}
		\label{tab:industry wage factors}
		\begin{tabular} {c c c c c c c c c c c c c c c c c c c} \toprule
			B & C & D & E & F & G & H & I & J & K & L & M & N & O & P & Q & R & S \\
			\toprule
			1.5 & 0.9 & 1.1 & 1 & 0.95 & 0.75 & 0.7 & 1 & 1.25 & 1.2 & 0.9 & 1.2 & 0.9 & 1.1 & 1.1 & 1 & 0.9 & 0.75 \\
			\bottomrule
		\end{tabular}
		\begin{minipage}{16cm}
			\vspace{0.1cm}
			\small Note: Factors generated using data from \textit{Average Weekly Earnings, Australia}, November 2021 cycle (\cite{australianbureauofstatistics_2022a})
		\end{minipage}
	\end{center}
\end{table} 

Table \ref{tab:error variance} contains the variance parameter values used in each size group for generating the wages data item.  These parameter values were chosen heuristically to obtain a realistic looking distribution for the resulting wages data within each size group.

\begin{table}[H]
	\begin{center}
		\caption{Variance terms for wages data item}
		\label{tab:error variance}
		\begin{tabular} {c c c} \toprule
			Size Group & Wage variance factor $(v_e)$ & Error variance $(v_s)$ \\
			\toprule
			0-4     & 0.10 & 50 \\ 
			5-19    & 0.10 & 75 \\
			20-49   & 0.15 & 100 \\
			50-99   & 0.15 & 150 \\
			100-149 & 0.15 & 150 \\
			150-199 & 0.15 & 150 \\
			200-249 & 0.15 & 200 \\
			249-299 & 0.20 & 250 \\
			300-349 & 0.20 & 250 \\
			349-399 & 0.20 & 250 \\
			400-449 & 0.20 & 250 \\
			449-499 & 0.20 & 250 \\
			500-999 & 0.20 & 275 \\
			1000+   & 0.15 & 325 \\
			\bottomrule
		\end{tabular}
	\end{center}
\end{table} 
\end{appendices}

\section*{ABS DataLab Disclaimer}

The parameters used to create the artificial population for the simulation study are based, in part, on data supplied to the ABS under the Taxation Administration Act 1953, A New Tax System (Australian Business Number) Act 1999, Australian Border Force Act 2015, Social Security (Administration) Act 1999, A New Tax System (Family Assistance) (Administration) Act 1999, Paid Parental Leave Act 2010 and/or the Student Assistance Act 1973. Such data may only used for the purpose of administering the Census and Statistics Act 1905 or performance of functions of the ABS as set out in section 6 of the Australian Bureau of Statistics Act 1975. No individual information collected under the Census and Statistics Act 1905 is provided back to custodians for administrative or regulatory purposes. Any discussion of data limitations or weaknesses is in the context of using the data for statistical purposes and is not related to the ability of the data to support the Australian Taxation Office, Australian Business Register, Department of Social Services and/or Department of Home Affairs’ core operational requirements.

Legislative requirements to ensure privacy and secrecy of these data have been followed. For access to PLIDA and/or BLADE data under Section 16A of the ABS Act 1975 or enabled by section 15 of the Census and Statistics (Information Release and Access) Determination 2018, source data are de-identified and so data about specific individuals has not been viewed in conducting this analysis. In accordance with the Census and Statistics Act 1905, results have been treated where necessary to ensure that they are not likely to enable identification of a particular person or organisation.

\newpage

\printbibliography

\end{document}